\documentclass[10pt,journal,twocolumn,romanappendices]{IEEEtran}
\IEEEoverridecommandlockouts
\usepackage{amsmath,amsthm,relsize,flushend}
\usepackage{amsfonts, amssymb, cuted}
\usepackage{cleveref}
\usepackage{mathtools}
\usepackage{algorithm}
\usepackage{cite}
\usepackage{comment}
\usepackage{xcolor}
\usepackage{soul}
\usepackage{graphicx}
\usepackage{tikz}
\usepackage{pgfplotstable}
\usepackage{pgfplots}
\usepackage{nicefrac}
\usepackage{enumitem}
\usepackage{algpseudocode}
\usepackage{amssymb}
\usepackage{algorithm}
\setlist[itemize]{topsep=0pt, parsep=0pt, partopsep=0pt, leftmargin=*}
\usepackage{subcaption}
\usepackage{bbm,microtype,setspace,url,caption}
\usepackage[font=footnotesize]{caption}
\usepackage{multirow}
\pgfplotsset{compat=1.15}
\usepackage{relsize}
\usetikzlibrary{patterns}
\usepackage{acro}
\usepackage{todonotes} 
\theoremstyle{plain}
\newtheorem{proposition}{Proposition}
\usepackage[algo2e]{algorithm2e} %

\renewcommand{\b}{\mathbf{b}}
\renewcommand{\c}{\mathbf{c}}

\newcommand{\g}{\mathbf{g}}

\newcommand{\p}{\mathbf{p}}

\newcommand{\w}{\mathbf{w}}

\newcommand{\y}{\mathbf{y}}
\newcommand{\z}{\mathbf{z}}

\newcommand{\0}{\mathbf{0}}

\newcommand{\B}{\mathbf{B}}

\newcommand{\G}{\mathbf{G}}
\renewcommand{\H}{\mathbf{H}}
\newcommand{\I}{\mathbf{I}}

\renewcommand{\P}{\mathbf{P}}

\newcommand{\U}{\mathbf{U}}

\newcommand{\X}{\mathbf{X}}
\newcommand{\Y}{\mathbf{Y}}
\newcommand{\Z}{\mathbf{Z}}

\newcommand{\setB}{\mathcal{B}}
\newcommand{\setC}{\mathcal{C}}

\newcommand{\setK}{\mathcal{K}}
\newcommand{\setL}{\mathcal{L}}
\newcommand{\setM}{\mathcal{M}}
\newcommand{\setN}{\mathcal{N}}

\newcommand{\setS}{\mathcal{S}}

\newcommand{\Compl}{\mbox{$\mathbb{C}$}}

\newcommand{\Diag}{\mathrm{Diag}}

\newcommand{\Exp}{\mathbb{E}}
\newcommand{\herm}{\mathrm{H}}

\renewcommand{\Pr}{\mathbb{P}}

\renewcommand{\Re}{\mathrm{Re}}

\newcommand{\tran}{\mathrm{T}}

\usepackage{etoolbox}
\AtBeginEnvironment{equation}{\setlength{\abovedisplayskip}{3pt}\setlength{\belowdisplayskip}{3pt}}
\AtBeginEnvironment{align}{\setlength{\abovedisplayskip}{3pt}\setlength{\belowdisplayskip}{3pt}}
\usepackage{amsthm}

\theoremstyle{remark}
\newtheorem{remark}{Remark}

\newcommand{\mse}{\mathrm{MSE}}
\newcommand{\sinr}{\mathrm{SINR}}

\newcommand{\ue}{\textnormal{\tiny{UE}}}

\newcommand{\beam}{\textnormal{\tiny{B}}}
\newcommand{\pre}{\textnormal{\tiny{P}}}

\newcommand{\Quant}{\mathrm{Q}}

\title{Codebook-Based Effective CSI Feedback for Precoding Design in MIMO Systems }
\author{
Bikshapathi Gouda,
Antti Arvola,~\IEEEmembership{Student Member,~IEEE},
Juha Karjalainen, 
Sami Hakola, 
and Antti T\"olli,~\IEEEmembership{Senior Member,~IEEE}
\thanks{Part of this work is presented at IEEE EuCN conference 2026~\cite{Gou26}.}
\thanks{Bikshapathi Gouda is with Skylo Technologies Europe Oy, Espoo, Finland (e-mail: bikshapathi@skylo.tech). This work was carried out while he was with the Centre for Wireless Communications, University of Oulu, Finland.}
\thanks{Antti Arvola and Antti T\"olli are with the Centre for Wireless Communications, University of Oulu, Finland (e-mail: \{antti.arvola, antti.tolli\}@oulu.fi).}
\thanks{Juha Karjalainen and Sami Hakola are with Technology Standards at Nokia, Oulu, Finland (e-mail: \{juha.p1.karjalainen, sami.hakola\}@nokia.com).}
\vspace{-6mm}}
\begin{document}
\maketitle

\begin{abstract}
Downlink multi-user multiple-input multiple-output (MIMO) precoding with limited channel state information (CSI) feedback in multi-cell systems is studied. Conventional codebook-based CSI feedback compresses the UE-specific physical channels, which leaves an inherent mismatch between the base-station (BS) precoders and the interference-aware UE combiners.  To address this limitation, an effective CSI (ECSI) feedback framework is proposed in which UEs compute their linear combiners from precoded downlink pilots, construct corresponding post-combining effective channels, and report compact quantized representations using the same payload structure as in conventional schemes. The BS reconstructs the effective CSI and iteratively refines its precoders through over-the-air signaling without incurring additional downlink overhead. Analytical expressions are derived to characterize the effective-channel estimation error under both conventional CSI and ECSI feedback, explicitly capturing the impact of beam selection and rank truncation. Simulations demonstrate fast convergence within a few iterations and up to 30\% average sum-rate improvement over conventional CSI-based precoding, with gains reaching 100\% at the 10th percentile of per-stream rates in interference-limited conditions.

\end{abstract}

\section{Introduction} \label{sec:INTRO}
The upcoming 6G systems extend previous cellular generations by further increasing the use of massive multiple-input multiple-output (MIMO) with large antenna arrays at both transmitter and receiver \cite{Raj20,dahlman20book,Mar10,Bjornson17MassiveMIMOBook}. Larger arrays provide higher spatial resolution, improved spatial multiplexing capability, and stronger beamforming gains; however, these benefits rely critically on sufficiently accurate channel state information (CSI). To support reliable CSI acquisition, recent 3GPP standards provide flexible pilot and reference signal (RS) configurations for channel sounding \cite{TS38211}. In time-division duplexing (TDD) systems, the base station (BS) can ideally obtain downlink CSI by exploiting channel reciprocity, since uplink and downlink share the same frequency band and experience the same physical propagation channel \cite{Mar10}. Accordingly, the BS may estimate the downlink channel from uplink pilot transmissions, such as sounding reference signals (SRS) transmitted by the user equipment (UE), provided that the SRS quality is sufficiently high, otherwise CSI-RS-based methods can be employed.

Extensive prior work has studied uplink pilot-based acquisition of downlink CSI in TDD systems, including reciprocity-based channel estimation and iterative transceiver optimization \cite{Shi11WMMSE}. Of particular relevance are over-the-air (OTA) bi-directional beamformer training schemes, in which BS precoders and UE combiners are iteratively refined using alternating downlink and uplink pilot signaling \cite{kom13,jay18,tol19,Shi14}. In these approaches, the BS first transmits precoded downlink pilots (e.g., CSI-RS or demodulation reference signals (DMRS)), enabling the UE to estimate the effective downlink channel and update the receive combiner. Subsequently, the UE transmits the precoded uplink pilots, based on which the BS updates its downlink precoder. This alternating procedure continues until convergence in mean-square error (MSE) or sum rate is achieved \cite{Shi11WMMSE, Scu14}. Similar OTA training principles have been extended to coordinated and cell-free massive MIMO systems to enable efficient beamformers \cite{kom13, Ngo17, kal18, Atz21,Gou24,gou24uldl}.

In practice, the reciprocity-based acquisition of downlink CSI is not always reliable. Limited SRS quality, hardware impairments, calibration errors, and asymmetric uplink and downlink capabilities at the UE may limit the validity of reciprocity. Uplink and downlink transmissions can rely on different radio-frequency (RF) chains or antenna subsets, and uplink transmit power limitations may result in low signal to noise ratio pilot reception at the BS \cite{Bjornson17MassiveMIMOBook,Vie17Reciprocity,dahlman20book,Mar10}. Moreover, practical constraints may restrict SRS configurations or periodicity. Furthermore, in frequency-division duplexing (FDD) systems, the channel is generally not reciprocal, since the uplink and downlink transmissions occupy different frequency bands~\cite{dahlman20book}. These limitations motivate explicit downlink CSI feedback mechanisms. As antenna array dimensions increase, direct feedback of the full channel matrix or its covariance becomes infeasible due to prohibitive overhead \cite{Love08overview,Jindal06TIT,Mon06,Love04ValueFeedback}. Consequently, a substantial body of literature has investigated structured CSI compression and limited-feedback techniques. Classical approaches include Grassmannian, random vector quantization codebooks, and channel-adaptive quantization \cite{Love08overview,Jindal06TIT,Mon06}. Compressive sensing methods can reduce the feedback by leveraging the angular and delay-domain sparsity in massive MIMO channels \cite{Gao15CSIT}. More recently, data-driven approaches have emerged, where deep-learning-based CSI feedback frameworks learn compact nonlinear representations of channel matrices while maintaining reconstruction fidelity \cite{Wen18DeepCSI,Guo20DLCSI}. Extensions incorporate recurrent and convolutional architectures to exploit temporal and frequency correlation, as well as model-driven and attention-based architectures for wideband CSI compression \cite{CsiNetLSTM,Song22TemporalCSI,Ju23,Ma25}.

The aforementioned works address general CSI compression, the structural design and evolution of standardized 3GPP codebooks have primarily been driven by industrial standardization rather than academic optimization. As highlighted in \cite{ning26}, comparatively limited academic effort has been devoted to systematic enhancement or comparative evaluation of NR codebook structures. In practice, most systems adopt the standardized Type-II codebook without iterative effective-channel refinement, which restricts available one-to-one comparison frameworks beyond the conventional Type-II baseline. In practical deployments, 3GPP New Radio (NR) adopts structured codebook-based CSI feedback, including Type-I and Type-II designs \cite{TS38214}. Type-I feedback provides a coarse representation through a precoder matrix indicator, rank indicator, and channel quality indicators. Type-II feedback offers higher spatial resolution by representing the estimated downlink channel in the beam domain. Specifically, the UE projects the channel onto a predefined wideband beam set, typically based on discrete Fourier transform (DFT) or oversampled DFT structures, selects a limited number of dominant beams, and reports their indices together with beam-specific and subband-specific quantized complex weighting coefficients \cite{TS38214,ning26,fu23}. Additional compression across frequency and time domains further reduces feedback overhead. These mechanisms enable reconstruction of a frequency-selective approximation of the physical downlink channel under strict payload constraints.

Despite its practical efficiency, conventional Type-II CSI feedback is designed to approximate the physical downlink channel and does not explicitly account for the receive combining strategy or instantaneous interference conditions at the UE. In multi-cell and interference-limited networks, coordinated transmission performance is highly sensitive to CSI accuracy and interference structure \cite{Gesbert10JSAC,Irmer11CoMP}. UEs typically employ interference-aware combiners, such as minimum MSE (MMSE) combining, whose optimality depends on the interference covariance in addition to the desired channel \cite{Bjornson12CombiningVsMultiplexing}. Consequently, when the BS reconstructs only the physical channel while the UE applies interference-dependent combining, a mismatch may arise between the BS-side channel representation and the post-combining effective channel experienced at the UE, potentially limiting interference suppression and cell-edge performance. These considerations motivate CSI feedback mechanisms that better capture the effective channel rather than solely the physical channel.

\smallskip
\noindent\textit{Contribution.} 
To address the mismatch between physical-channel-based feedback and interference-aware combining, we propose an \textit{effective CSI (ECSI)} feedback framework in which UEs report compressed representations of their post-combining effective channels while preserving compatibility with the Type-II codebook payload structure. By jointly leveraging cell-specific CSI-RS and  UE-specific precoded CSI-RS or DMRS, the BS enables OTA refinement of precoders through iterative effective-channel updates. In contrast to conventional Type-II feedback, which compresses a rank-reduced beam-domain representation of the physical channel and may incur non-negligible rank-truncation mismatch, the proposed ECSI framework directly compresses the post-combining effective channel, thereby eliminating the rank-truncation error component and providing a closer approximation to the true effective CSI under the same feedback budget. We develop a unified analytical characterization of the effective-channel estimation error for both conventional CSI and ECSI feedback, explicitly decomposing the error into beam-selection, rank-truncation, and AWGN-induced components and revealing the structural advantage of ECSI in interference-limited regimes. The framework is further extended to coordinated multi-cell beamforming scenarios, where, in addition to the effective CSI of the serving BS, the UEs feed back the ECSI of interfering BSs to enable interference-aware precoding across cells. 

\smallskip

\noindent\textit{Main contributions are summarized as follows.}
\begin{itemize}
\item An ECSI feedback framework is proposed that preserves compatibility with the Type-II codebook while enabling direct compression of the post-combining effective channel for improved precoder--combiner alignment.
\item A unified analytical characterization of the effective-channel estimation error is derived for both conventional CSI and ECSI feedback, decomposing it into beam-selection, rank-truncation, and AWGN-induced components.
\item It is shown that conventional Type-II CSI feedback incurs effective-channel mismatch due to rank truncation, whereas the proposed ECSI framework eliminates this component.
\item The framework is extended to coordinated multi-cell precoding by incorporating effective CSI feedback of interfering links for interference-aware beamforming.
\item Numerical results demonstrate fast OTA convergence and consistent sum-rate gains, particularly in interference-limited and cell-edge regimes.
\end{itemize}

\smallskip

\noindent\textit{Outline.}
The remainder of the paper is organized as follows. 
Section~\ref{sec:SM} introduces the system model and the MSE-based beamforming design. Section~\ref{sec:CSI_comp} describes the conventional CSI compression and feedback framework. 
Section~\ref{sec:ECSI} presents the proposed effective CSI framework and its analytical characterization. 
Section~\ref{sec:NUM} provides numerical results, and Section~\ref{sec:CONC} concludes the paper.

\smallskip
\noindent\textit{Notation.}
Lowercase and uppercase boldface letters denote vectors and matrices, respectively. $(\cdot)^{\tran}$ and $(\cdot)^{\herm}$ represent the transpose and
Hermitian transpose operators, respectively. $|\cdot|$ denotes the absolute value
of a scalar, while $|\mathcal{S}|$ denotes the cardinality of the set $\mathcal{S}$.
$\bar{\mathcal S}$ denotes the complement of the set $\mathcal S$ with respect to
its ambient index set. The notation $\|\cdot\|$ and $\|\cdot\|_{\mathrm F}$ denote
the Euclidean and Frobenius norms, respectively. $\Re[\cdot]$, $\Exp[\cdot]$, and
$\Pr(\cdot)$ denote the real-part, expectation, and probability operators,
respectively. $\mathbf{I}_{A}$ denotes the $A$-dimensional identity matrix.
$\Diag(\cdot)$ produces a diagonal matrix with the elements of a vector argument
or the diagonal elements of a square matrix argument on its diagonal.
$[a_{1}, \ldots, a_{L}]$ denotes horizontal concatenation, whereas
$\{a_{1}, \ldots, a_{L}\}$ and $\{ a_{\ell} \}_{\ell \in \setL}$ denote sets; the
latter notation is occasionally relaxed as $\{ a_{\ell} \}$ for brevity.
$\mathcal{CN}(0, \sigma^{2})$ denotes the circularly symmetric complex Gaussian
distribution with zero mean and variance $\sigma^{2}$.

\section{System Model and Beamforming Design} \label{sec:SM}
We consider a multi-cell downlink system with a set of BSs~$\setB=\{1,\dots,B\}$ equipped with $M$ antennas each, serving a set of UEs $\setK=\{1,\dots,K\}$. Each UE~$k$ is equipped with $N$ antennas and is associated with a unique serving BS denoted by $b_k \in \setB$, where $\setK_b = \{k \in \setK : b_k = b\}$ denotes the set of UEs served by BS~$b$. Each UE $k$ is assigned a set of data streams $\setS_k=\{1,\ldots,S_k\}$, and the downlink channel between BS~$b$ and UE~$k$ is denoted by $\H_{b,k} \in \Compl^{N\times M}$. Without loss of generality, we assume a narrowband flat-fading channel model corresponding to a single subband. To this end, the serving BS~$b_k$ applies a linear precoder $\w_{k,s}\in\Compl^{M\times 1}$ to transmit the data symbol $d_{k,s}\sim\setC\setN(0,1)$ intended for stream $s \in \setS_k$. Consequently, the received signal at UE~$k$ is given by
\begin{align}\label{eq:y_k}
\y_k =  
\sum_{\bar b \in \setB} \H_{\bar b,k} 
\sum_{j\in\setK_{\bar b}}\sum_{s\in\setS_j}\w_{j,s}\, d_{j,s} 
+ \z_k 
\in \Compl^{N\times 1},
\end{align}
where $\z_k\sim\setC\setN(\0,\sigma_{\ue}^2 \I_N)$ is the additive white Gaussian noise (AWGN). Furthermore, to estimate the transmitted symbol $d_{k,s}$, UE~$k$ applies a linear combiner $\c_{k,s}\in\Compl^{N\times 1}$, which yields the soft estimate
\begin{align}\label{eq:dhat}
\hat d_{k,s} = \c_{k,s}^{\herm}\y_k.
\end{align}
Then, the corresponding SINR for the stream $s$ of UE~$k$ is given in~\eqref{eq:SINR_s_k}.
\begin{figure*}[t!]
\begin{align}\label{eq:SINR_s_k}
\sinr_{k,s} &
\triangleq \frac{\big| \c_{k,s}^{\herm}\H_{b_k,k}\w_{k,s} \big|^2}
{ \sum_{{\bar b}} \sum_{{j\in\setK_{\bar b}}} \sum_{{\bar s\in\setS_j}} \big| \c_{k,s}^{\herm}\H_{\bar b,k}\w_{{j},\bar{s}} \big|^2 - \big| \c_{k,s}^{\herm}\H_{b_k,k}\w_{k,s} \big|^2
+ \sigma_{\ue}^2 \|\c_{k,s}\|^2 }
\end{align}
\hrule
\end{figure*}
Finally, the sum rate (bps/Hz) across all UEs and streams is
\begin{align}\label{eq:sumrate}
R_{\text{sum}} =  \sum_{k\in\setK}  \sum_{s\in\setS_k}\log_2\big(1+\sinr_{k,s}\big),
\end{align}
which serves as an upper bound on the achievable rate and will be used to evaluate the performance of the system in Section~\ref{sec:NUM}. In the following, we discuss the designing of precoders and combiners based on the MSE criterion.

\subsection{Beamforming Design Based on MSE} \label{sec:MSEJDU}
The MSE of the estimated symbol for stream~$s$ of UE~$k$ is defined as  
\begin{align}
\mse_{k,s} 
&= \Exp \!\left[ \big|\c_{k,s}^{\herm}\y_{k} - d_{k,s}\big|^2 \right] \label{eq:MSE_s_k_2} \\[0.5ex]
&=
\sum_{\bar b\in\mathcal{B}}
\sum_{j\in\mathcal{K}_{\bar b}}
\sum_{\bar s\in\mathcal{S}_j}
\big|
\c_{k,s}^{\herm}\H_{\bar b,k}\w_{j,\bar s}
\big|^2  \nonumber \\
&\quad 
- 2\Re\!\big[
\c_{k,s}^{\herm}\H_{b_k,k}\w_{k,s}
\big]  + \sigma_{\ue}^2\|\c_{k,s}\|^2 + 1.
\end{align}
The above MSE is not jointly convex with respect to $\w_{k,s}$ and $\c_{k,s}$. Therefore, we adopt an alternating optimization approach, i.e., for the fixed set of precoders $\{\w_{k,s}\}$, the combiners $\{\c_{k,s}\}$ are updated, and vice versa. In the following, we derive the corresponding updates for the precoders and combiners under the BS power constraint.

\textbf{Combiner update.}
 For a fixed set of precoders $\{ \w_{k,s} \}$, the combiner for the stream $s$ of UE~$k$ is obtained by minimizing the MSE in~\eqref{eq:MSE_s_k_2} with respect to $\c_{k,s}$, which yields the MMSE combiner as given in~\eqref{eq:uebf_MB_2}.
\begin{figure*}
\begin{align}
\c_{k,s}
& =
\Bigg(\sum_{ j \in\mathcal{K}_{ b_k}}
\sum_{\bar s\in\mathcal{S}_j}
\H_{ b_k,k}\w_{{j},\bar s}
\w_{{j},\bar s}^{\herm}\H_{b_k,k}^{\herm}
+ \underbrace{\sum_{\bar b\in\mathcal{B} \setminus \{b_k\}}
\sum_{j\in\mathcal{K}_{\bar b}}
\sum_{\bar s\in\mathcal{S}_j}
\H_{\bar b,k}\w_{{j},\bar s}
\w_{{j},\bar s}^{\herm}\H_{\bar b,k}^{\herm}}_{\text{inter-cell term}} +
\sigma_{\ue}^{2}\I_N
\Bigg)^{-1} \H_{b_k,k}\w_{k,s} \label{eq:uebf_MB_2}
\end{align}
\hrule
\end{figure*}

\textbf{Precoder update.}
 For the fixed set of combiners $\{ \c_{k,s} \}$, the precoder for the stream $s$ of UE~$k$ is obtained by minimizing the sum of the MSEs in~\eqref{eq:MSE_s_k_2} of all the UEs and streams with respect to $\w_{k,s}$, subject to the BS~$b$ power constraint $\rho_b$, i.e.,
\begin{align}
\begin{array}{cl}
\displaystyle\underset{\{\w_{k,s}\}}{\mathrm{minimize}}
& \displaystyle
\sum_{b\in\mathcal{B}}
\sum_{k\in\mathcal{K}_b}
\sum_{s\in\mathcal{S}_k}
\mse_{k,s} \\[1.5ex]
\mathrm{subject~to}
& \displaystyle
\sum_{k\in\mathcal{K}_b}
\sum_{s\in\mathcal{S}_k}
\|\w_{k,s}\|^{2}\le\rho_b,
\quad \forall b\in\mathcal{B}.
\end{array}
\label{eq:MSE_precoder_MB_problem}
\end{align}

The Karush-Kuhn-Tucker (KKT) conditions of~\eqref{eq:MSE_precoder_MB_problem} yield the MMSE precoder given in~\eqref{eq:bsbf_MB_2}.
\begin{figure*}
\begin{align}
\w_{k,s}
& =
\Bigg(
\sum_{ j\in\mathcal{K}_{ b_k}}
\sum_{\bar s\in\mathcal{S}_j}
\H_{ b_k,j}^{\herm}\c_{{j},\bar s}
\c_{{j},\bar s}^{\herm}\H_{b_k,j}
+ \underbrace{\sum_{\bar b\in\mathcal{B} \setminus \{b_k\}}
\sum_{ j\in\mathcal{K}_{\bar b}}
\sum_{\bar s\in\mathcal{S}_j}
\H_{ b_k,j}^{\herm}\c_{{j},\bar s}
\c_{{j},\bar s}^{\herm}\H_{ b_k,j}
+}_{\text{inter-cell terms}}
\nu_{b}\I_M
\Bigg)^{-1}  \H_{b_k,k}^{\herm}\c_{k,s}
\label{eq:bsbf_MB_2}
\end{align}
\hrule
\end{figure*}
Here, $\nu_{b} \ge 0$ is the Lagrangian dual variable corresponding to the BS-$b$ power constraint. In the following, we discuss the OTA pilot signaling required for the practical design of the UE combiners and the BS precoders.

\subsection{OTA Pilot Signaling} 

\textit{\textbf{UE-specific pilot signaling.}}   During the connected mode, for a fixed set of precoders $\{\w_{k,s}\}$, the BSs simultaneously transmit UE-specific CSI-RS or DMRS. Using the received UE-specific CSI-RS or DMRS, each UE can estimate the post-precoding effective channel and accordingly compute or update its linear combiner. Let $\p_{k,s}\in\Compl^{\tau^{\pre}\times 1}$ denote the pilot sequence associated with $\w_{k,s}$, normalized such that $\|\p_{k,s}\|^2=\tau^{\pre}$, and pairwise orthogonal across all the streams and BSs, i.e., $\p_{k,s}^{\herm}\p_{j,\bar s}=0$ for $(k,s)\neq(j,\bar s)$. Accordingly, the transmitted signal of BS~$b$ is
\begin{align}
\label{eq:X_b_dl}
\X_b^{\pre} = \sum_{k\in\setK_b}\sum_{s\in\setS_k} \w_{k,s}\,\p_{k,s}^{\herm} \;\in\; \Compl^{M\times \tau^{\pre}}.
\end{align}
The corresponding received signal at UE~$k$ is given as
\begin{align}\label{eq:Y_k_dl_2}
\Y_{k}^{\pre} & = \sum_{\bar b \in \setB} \X_{\bar b}^{\pre}+ \Z_{k}^{\pre} \;\in\; \Compl^{N\times \tau^{\pre}} \\
& = \sum_{\bar b \in \setB} \sum_{j \in \setK_{\bar b}} \sum_{\bar s \in \setS_j} 
\H_{\bar b,k} \w_{j,\bar s}  \p_{j,\bar s}^{\herm} + \Z_{k}^{\pre},
\end{align}
where $\Z_{k}^{\pre}$ is AWGN. Using the received signal, UE~$k$ can compute its combiner as
\begin{align}\label{eq:rxmmse_est}
\c_{k,s} = \big(\Y_{k}^{\pre} (\Y_{k}^{\pre})^{\herm}\big)^{-1} \Y_{k}^{\pre} \p_{k,s},
\end{align}
which simplifies to~\eqref{eq:uebf_MB_2} as ${\tau^{\pre}}\rightarrow \infty$ or at high SINR.

\textit{\textbf{Cell-specific pilot signaling.}}The BSs can transmit the cell-specific CSI-RS, to enable the UEs to estimate the physical downlink channels and report CSI for precoder design. Specifically, each UE first estimates the channel based on the received CSI-RS, after which the estimated channel is compressed to reduce feedback overhead, and the compressed CSI is then fed back to the BS. To improve compression efficiency, the CSI is typically represented in the beam domain by exploiting the limited angular spread around each UE. In NR systems, this can be realized through different CSI-RS configurations, such as conventional antenna-port-based CSI-RS or beamformed CSI-RS, the latter typically used with port-selection (PS) codebooks~\cite{TS38211}. In the AP-based approach, the UE estimates the antenna-domain channel and performs the transformation to the beam domain prior to feedback, whereas in the beam-based (PS) approach, the UE directly estimates the beam-domain channel. In this work, we adopt the beam-based CSI-RS transmission, although the proposed framework readily extends to the antenna-port-based case. To this end, we assume that each BS employs orthogonal time or frequency resources when transmitting beam-specific pilots. For any BS, let us define $\setM \triangleq \{1,\ldots,M\}$, and let $\b_m \in \mathbb{C}^{M\times 1}$ denote the $m$-th beam (column) of a unitary DFT codebook with $\B=[\b_1,\ldots,\b_M]\in\mathbb{C}^{M\times M}$. Although in practice the beam domain may be constrained to a subset of beams, here we assume that the full $M\times M$ DFT codebook is available to simplify the analysis; all subsequent developments extend directly to reduced or structured beam sets.

Let $\tilde{\p}_{\bar b,m}\in\mathbb{C}^{\tau^{\beam}\times 1}$ be the pilot sequence
assigned to beam $m$ of BS~$\bar b$, satisfying $\|\tilde{\p}_{\bar b,m}\|^2=\tau^{\beam}$ and
$\tilde{\p}_{\bar b,m}^{\herm}\tilde{\p}_{\bar b,\bar m}=0$ for $m\neq \bar m$. Then, BS~$\bar b$
transmits pilots across all beams as
\begin{align}
\label{eq:X_b_dl_p}
\X_{\bar b}^{\beam} \triangleq \sum_{m\in\setM} \b_{\bar b,m} \tilde \p_{\bar b,m}^{\herm}
= \B_{\bar b}\tilde \P_{\bar b}^{\herm} \in \Compl^{M\times \tau^{\beam}},
\end{align}
where $\tilde \P_{\bar b}=[\tilde \p_{\bar b,1},\ldots,\tilde \p_{\bar b,M}]\in\Compl^{\tau^{\beam}\times M}$. The received pilot signal at UE $k$ is 
\begin{align}
\label{eq:Y_k_dl}
\Y_{k}^{\beam} & \triangleq \sum_{\bar b \in \setB} \H_{\bar b,k} \X_{\bar b}^{\beam} + \Z_{k}^{\beam} \in \Compl^{N\times \tau^{\beam}} \\
& = \sum_{\bar b \in \setB} \H_{\bar b,k} \B_{\bar b} \tilde \P_{\bar b}^{\herm} + \Z_{k}^{\beam},
\end{align}
where $\Z_{k}^{\beam}$ denotes AWGN. From $\Y_{k}^{\beam}$, UE~$k$ can obtain channel information of all the BSs. Consequently, UE~$k$ estimates the beam domain channel $\G_{\bar b,k}\triangleq\H_{\bar b, k}\B_{\bar b}\in\Compl^{N\times M}$ of BS~$\bar b$ from \eqref{eq:Y_k_dl}, using the least-squares (LS) estimator as follows 
\begin{align}\label{eq:bd_ch} 
\hat{\G}_{\bar b, k} & = \frac{1}{\tau^{\beam}}\,\Y_{k}^{\beam} \tilde \P_{\bar b} \\ 
& = \G_{\bar b, k} + \frac{1}{\tau^{\beam}}\,\Z_{k}^{\beam} \tilde \P_{\bar b}. 
\end{align}
The above estimate simplifies to $\G_{\bar b, k}$ as ${\tau^{\beam}} \rightarrow \infty$ or at high SINR.  This information is used to perform CSI compression and feedback for precoder design, as detailed in Section~\ref{sec:CSI_comp}.

\section{Baseline Framework for CSI Compression and Precoding Design} \label{sec:CSI_comp}
To reduce channel feedback overhead, 3GPP adopts codebook-based CSI reporting, where the channel is first projected onto a predefined beam domain and subsequently compressed via rank reduction based on dominant eigenmodes. The CSI feedback can be further compressed over time by exploiting temporal correlation using Doppler information, and in wideband systems by leveraging sub-band correlation in the frequency domain. Without loss of generality, we focus on a fixed time instant and a single sub-band, noting that additional compression across Doppler and delay domains can be straightforwardly incorporated following the same principles~\cite{ning26, fu23}.

This beam-domain compression and rank reduction procedure follows the state-of-the-art CSI reporting framework specified in 3GPP NR, as described in Release~17 and further refined in Release~18/19~\cite{TS38211,ning26,fu23}. While the CSI compression and feedback discussed below can apply to a general channel realization, including potential interfering links, the current 3GPP specifications define this CSI reporting procedure only for the serving cell, or for all active transmission points in the case of coherent joint transmission.

\subsection{CSI Compression at the UE}\label{subsec:csi_comp}
In this section, we discuss CSI compression at the UE using beam selection and rank reduction, as well as the quantization of the CSI feedback and the reconstruction of the reduced channel matrix at the BS. In the following, we discuss the selection of beams from the estimated beam domain channel $\hat{\G}_{\bar b, k}$ given in \eqref{eq:bd_ch}.

\textbf{\textit{Beam selection.}}
Let $\G_{\bar b,k} = [\,\g_{\bar b,k,1},\ldots,\g_{\bar b,k,M}\,]$, where $\g_{\bar b, k,m} \in\Compl^{N\times 1}$ is the effective channel when BS~$\bar b$ transmits on beam $\b_{\bar b,m}$. From the LS estimate of $\G_{\bar b,k}$, i.e., $\hat{\G}_{\bar b, k} = [\,\hat{\g}_{\bar b,k,1},\ldots,\hat {\g}_{\bar b,k,M}\,] $, UE $k$ computes the per-beam energy metric $\hat \alpha_{\bar b,k,m}=\|\hat{\g}_{\bar b,k,m}\|_2^2$ and selects the $L$ largest values~\cite{TS38214}. Let $\mathcal{I}_{\bar b, k}\subseteq\{1,\ldots,M\}$ be the index set of the selected beams with $|\mathcal{I}_{\bar b, k}|=L$. Then, the reduced beam-domain channel estimate is
\begin{align}\label{eq:bm_ch}
\tilde{\G}_{\bar b, k} \triangleq \H_{\bar b, k} \B_{\mathcal{I}_{\bar b, k}} + \frac{1}{\tau^{\beam}}\,\Z_{k} \tilde \P_{\mathcal{I}_{\bar b, k}} \;\in\; \Compl^{N\times L}, 
\end{align}
where $\B_{\mathcal{I}_{\bar b, k}} \in \Compl^{M \times L}$ concatenation of the selected columns of $\B_{\bar b}$ and $\P_{\mathcal{I}_{\bar b, k}} \in \Compl^{M \times L}$ the corresponding pilot sequences. Furthermore, the beam-domain channel can be further compressed by rank reduction, using singular value decomposition (SVD), as discussed below.

\textit{\textbf{SVD and rank reduction.}}  
To further compress the beam-domain channel, the UE performs the SVD and retains only $R$ dominant singular modes. The SVD of beam-domain channel between UE~$k$ and BS~$\bar b$ can be written as
\begin{align}
\tilde{\mathbf{G}}_{\bar b, k} = \tilde{\mathbf{U}}_{\bar b, k}  \tilde {\boldsymbol{\Lambda}}_{\bar b, k}  \tilde {\mathbf{V}}_{\bar b, k} ^{\herm},
\end{align}
where $\tilde{\mathbf{U}}_{\bar b, k} \in \mathbb{C}^{N\times N}$, $\tilde{\mathbf{V}}_{\bar b, k} \in \mathbb{C}^{L\times L}$ are unitary matrices, and  $\tilde{\boldsymbol{\Lambda}}_{\bar b, k} \in \mathbb{C}^{N\times L}$ is a diagonal matrix containing the singular values, i.e., $\tilde{\boldsymbol{\Lambda}}_{\bar b, k}  \triangleq \Diag(\tilde{\lambda}_{\bar b, 1,k}, \dots, \tilde{\lambda}_{\bar b, \min( N,L),k})$. The UE selects $R \le \min(N, L)$ strongest singular values, along with their corresponding right singular vectors to compress the channel information~\cite{TS38214}. The compressed information is then quantized and fed back to the serving BS, as discussed below.

For each rank $r \in \{1,\dots,R\}$, the UE forms the normalized compressed channel coefficients by considering the maximum value for that rank, as follows:
\begin{align}
\bar v_{\bar b, l,r,k} = \frac{1}{a_{\bar b, r,k}} \, \lambda_{\bar b, r,k} \tilde v_{\bar b, l,r,k},
\end{align}
where $
a_{\bar b, r, k} =  \max_{l=1,\dots,L} \big| \tilde \lambda_{\bar b, r,k} \tilde v_{\bar b, l,r,k} \big|,$ and $\tilde v_{\bar b, l,r,k}$ denotes the $l$-th element of $r$-th right singular vector of UE~$k$ and BS~$\bar b$. Each coefficient $\bar v_{\bar b, l,r,k}$ is expressed in polar coordinate form as
\begin{align}
\bar v_{\bar b, l,r,k} = \big|\bar v_{\bar b, l,r,k}\big| \, e^{j\theta_{\bar b, l,r,k}},
\end{align}
with magnitude, i.e., $\big|\bar v_{\bar b, l,r,k}\big|$ quantized to $q_m$ levels, chosen from $ \{ 1, 1/\sqrt{2}, 1/\sqrt{4},  \dots, 1/\sqrt{2^{q_m-2}}, 0 \},$ and the phase, i.e., $\theta_{\bar b, l,r,k}$ is quantized using $q_p$ uniformly spaced levels over $[0,2\pi)$~\cite{TS38214}.  Finally, we define $\bar{\mathbf{V}}_{\bar b, k}^{(\Quant)} \in \mathbb{C}^{R \times L}$ as the quantized compressed channel between BS~$\bar b$ and UE~$k$, which is fed back to the serving BS~$b_k$ together with the normalized coefficients and the beam indices $\mathcal{I}_{\bar b, k}$.

\textit{\textbf{Channel reconstruction at the BS.}}  
The reported beam indices $\mathcal{I}_{\bar b, k}$ and the quantized compressed channel matrix $\bar{\mathbf{V}}_{\bar b, k}^{(Q)}$ are fed back to the serving BS. The serving BS~$b_k$ then forwards the compressed CSI to the respective BS~$\bar b$ via the backhaul link. The reconstructed reduced-dimension channel of UE~$k$ at BS~$\bar b$ is given as
\begin{align}\label{eq:com_rec_ch}
\bar{\mathbf{H}}_{\bar b, k} =  \bar{\mathbf{V}}_{\bar b, k}^{(\Quant)} \mathbf{B}_{\mathcal{I}_{\bar b, k}}^{\herm} \in \mathbb{C}^{R \times M}.
\end{align}
This reconstruction preserves the dominant channel subspace up to a unitary transformation arising from the omission of $\tilde{\mathbf{U}}_{\bar b, k}$, thereby enabling efficient beamforming and scheduling with reduced feedback.

\begin{algorithm}[t!]
\small
\begin{spacing}{1.1}
\textbf{Initialization:}
$L,R$, $q_m,q_p$, $\mathbf p_{k,s}$,
$\tilde{\mathbf p}_{\bar b,m}$, and
$\mathcal B_k$ for all UEs.

\begin{itemize}

\item Each BS~$\bar b$ transmits beam-specific pilots
$\tilde{\mathbf p}_{\bar b,m}$.

\item Each UE~$k$ estimates
$\{\tilde{\mathbf G}_{\bar b,k}\}_{\bar b\in\mathcal B_k}$
as in~\eqref{eq:bm_ch}, and reports
$\bar{\mathbf V}^{(Q)}_{\bar b,k}$ and
$\mathcal I_{\bar b,k}$.

\item The reduced-dimension channels
$\bar{\mathbf H}_{\bar b,k}\in\mathbb C^{R\times M}$,
$\forall\,\bar b\in\mathcal B_k$,
are reconstructed as in~\eqref{eq:com_rec_ch}
and made available for precoder design.

\item Initialize the virtual combiner
$\mathbf c_{k,s}\in\mathbb C^{R\times1}$.

\end{itemize}

\textbf{Repeat until convergence:}

\begin{enumerate}

\item Compute the precoder
$\mathbf w_{k,s}$
using~\eqref{eq:bsbf_MB_2}
by replacing
${\mathbf H}_{\bar b,k}$
with
$\bar{\mathbf H}_{\bar b,k}$.
Include the inter-cell terms if
$|\mathcal B_k|>1$.

\item Compute the combiner
$\mathbf c_{k,s}$
using~\eqref{eq:uebf_MB_2}
by replacing
${\mathbf H}_{\bar b,k}$
with
$\bar{\mathbf H}_{\bar b,k}$.
Include the inter-cell terms if
$|\mathcal B_k|>1$.

\end{enumerate}

\textbf{End}

\begin{itemize}

\item The converged precoder
$\mathbf w_{k,s}$
is used for data transmission.

\end{itemize}

\end{spacing}
\caption{Baseline beamforming design with CSI feedback and optional inter-cell interference}
\label{alg:csi_int}
\end{algorithm}

\subsection{Precoding Design at the BS}\label{sec:pre_csi}
In this section, we discuss the design of the BS precoders under different CSI availability scenarios. Let $\mathcal{B}_{k}$ denote the set of BSs whose CSI is available for UE~$k$. In the absence of inter-cell information, this set reduces to the serving BS only, i.e., $\mathcal{B}_{k} = \{b_k\}$ and $|\mathcal{B}_{k}| = 1$; otherwise, $|\mathcal{B}_{k}| > 1$, where the UE also feeds back the CSI of a subset of interfering BSs. Accordingly, the UE~$k$ measures the channels of all the BSs in $\mathcal{B}_{k}$, e.g., via orthogonal pilot resources assigned across all the BSs as in multi-TRP scenario~\cite{TS38214,ning26}, and feeds back the corresponding compressed CSI to its serving BS~$b_k$. The serving BS reconstructs the reduced-dimension channel $\bar{\mathbf{H}}_{\bar b,k} \in \mathbb{C}^{R \times M}$ for all $\bar b \in \mathcal{B}_{k}$ and forwards the non-serving CSI components to the respective BSs via backhaul links when $|\mathcal{B}_{k}| > 1$. Using the reconstructed channels, each BS~$\bar b \in \mathcal{B}_{k}$ iteratively computes the precoder $\mathbf{w}_{k,s} \in \mathbb{C}^{M \times 1}$ according to~\eqref{eq:bsbf_MB_2} and the reduced-dimension (virtual) combiner $\mathbf{c}_{k,s} \in \mathbb{C}^{R \times 1}$ according to~\eqref{eq:uebf_MB_2}, by replacing $\mathbf{H}_{\bar b,k}$ with $\bar{\mathbf{H}}_{\bar b,k}$.

If $|\mathcal{B}_{k}| = 1$, the design reduces to the single-cell case, where only the serving BS~$b_k$ is involved and inter-cell interference is ignored. Otherwise, when $|\mathcal{B}_{k}| > 1$, the inter-cell interference terms are explicitly incorporated into the precoder design. The iterative procedure is initialized with a randomly selected combiner $\mathbf{c}_{k,s}$, and the converged precoder $\mathbf{w}_{k,s}$ is adopted at the serving BS~$b_k$. Note that the combiner obtained during this iterative design serves only as an auxiliary variable for the precoder optimization. At the receiver, the UE applies the true combiner computed from its actual effective channel, and the corresponding true SINR is used in all performance evaluations. The complete procedure for obtaining the precoder is summarized in Algorithm~\ref{alg:csi_int}, which can be implemented in either a distributed or a centralized manner. In a distributed implementation, each BS computes its precoder locally using the available CSI together with the CSI and updated beamforming information exchanged with the cooperating BSs via backhaul links when $|\mathcal{B}_{k}| > 1$. In a centralized implementation, the BSs forward the available CSI to a central unit, which computes the precoders and combiners for all BSs and returns the resulting precoders to the corresponding BSs. In the following, we present the proposed ECSI framework, including effective channel compression, precoding design, and analytical performance characterization.

\section{ECSI Compression and Precoding Design} \label{sec:ECSI}

The conventional reduced-dimension CSI $\bar{\mathbf H}_{b_k,k}$ provides a compact representation of the physical channel, but does not explicitly account for the UE-side combining operation applied at the receiver. As a result, the legacy codebook-based CSI feedback framework designs the BS precoders based on a channel description that is decoupled from the actual interference-aware combining used at the UE, limiting its effectiveness in practical multi-stream and multi-UE scenarios with significant inter-stream, inter-UE, and inter-cell interference.

In this work, we propose an ECSI feedback scheme that aligns BS precoder design with the UE-side combining operation. Each UE operates in connected mode and utilizes UE-specific precoded CSI-RS or DMRS to estimate the downlink channels and compute interference-aware linear combiners from precoded pilots as in~\eqref{eq:rxmmse_est}. Based on these combiners, the UE feeds back its post-combining effective channel to the serving BS~$b_k$. Compared to conventional CSI feedback, the proposed approach inherently captures inter-stream, inter-UE, and inter-cell interference, thus improving precoder–combiner alignment. As in the baseline scheme, compression is performed in the angular (DFT beam) domain with $L$ dominant beams. The difference lies in the reduction of the receive dimension: the baseline applies an explicit rank reduction, projecting the channel onto its $R$ dominant eigen-directions, which is interference-agnostic. In the proposed scheme, this reduction is instead performed implicitly by the receive combiner, so the UE reports one $L$-dimensional beam-domain effective channel per stream and no additional rank-reduction processing is required. The effective channels can be further refined through OTA signaling to mitigate compression and quantization errors. In the following, we describe the computation of the ECSI at the UE and its beam-domain compression.

\subsection{ECSI Compression at the UE} \label{subsec:ecs_comp}
Building on the beam-domain channel estimate $\hat{\G}_{\bar b,k}$ obtained in Section~\ref{sec:CSI_comp}, and using the combiner $\mathbf{c}_{k,s}$ obtained from~\eqref{eq:rxmmse_est}, UE~$k$ computes the beam-domain effective channel for stream $s$ of BS~$\bar b$ as
\begin{align}\label{eq:esti_eff_beam}
    \hat{\mathbf{t}}_{\bar b,k,s} \triangleq \hat{\G}_{\bar b,k}^{\herm} \mathbf{c}_{k,s} \in \mathbb{C}^{M \times 1}.
\end{align}
The beam-domain transformation and the receive combining act on the transmit and receive dimensions of the channel, respectively, and are therefore interchangeable: combining followed by beam-domain compression yields the same effective channel as compression followed by combining. Using the beam-domain effective channels in~\eqref{eq:esti_eff_beam}, the UE performs beam selection as follows.

\textbf{\textit{Beam selection.}} 
For beam-domain compression, the UE selects a common set of dominant beams across all streams to reduce feedback overhead and ensure a consistent subspace representation. The received energy on beam $\mathbf{b}_{\bar b,m}$ is computed as
$\hat \beta_{\bar b, k, m} = \sum_{s \in \setS_k} | \hat{\mathbf g}_{\bar b, k,m}^{\herm}\mathbf{c}_{k,s} |^2,
$ where $\hat{\mathbf g}_{\bar b, k,m}$ denotes the $m$-th column of $\hat{\G}_{\bar b,k}$, as defined in Section~\ref{sec:CSI_comp}. The $L$ beams with the largest energy are then selected. Their indices form the set $\mathcal{I}_{\bar b, k} \subseteq \{1,\dots,M\}$ with $|\mathcal{I}_{\bar b, k}|=L$. Accordingly, substituting the LS estimate of the beam-domain channel into \eqref{eq:esti_eff_beam}, the reduced-dimension beam-domain effective channel is given as
\begin{align}\label{eq:beam_eff}
    \tilde{\mathbf{t}}_{\bar b, k,s} \triangleq \mathbf{B}_{\mathcal{I}_{\bar b, k}}^{\herm} \H_{\bar b, k}^{\herm}\mathbf{c}_{k,s}
    + \frac{1}{\tau^{\beam}}\,\P_{\mathcal{I}_{\bar b, k}}^{\herm} \Z_{k}^{\herm}  \mathbf{c}_{k,s}
    \in \mathbb{C}^{L \times 1},
\end{align}
where the second term represents the projected AWGN component after beam selection and combining. The entries of $\tilde{\mathbf{t}}_{\bar b, k,s}$ are quantized in amplitude and phase as described in Section~\ref{sec:CSI_comp}.  Finally, $\hat {\mathbf{t}}^{(Q)}_{\bar b,k,s}$ denotes the quantized compressed effective channel, which is fed back to the serving BS~$b_k$ together with normalization factors and beam indices $\mathcal{I}_{\bar b, k}$.

\textit{\textbf{Channel reconstruction at the BS.}}  
The serving BS~$b_k$ forwards the reported beam indices $\mathcal{I}_{\bar b, k}$ and the quantized ECSI vectors $\{\hat{\mathbf{t}}^{(Q)}_{\bar b, k,s}\}$ to the respective BSs. Based on this information, BS~$\bar b$ reconstructs the effective channel associated with stream~$s$ of UE~$k$ as
\begin{align}\label{eq:rec}
    \tilde{\mathbf{f}}_{\bar b, k,s} = \mathbf{B}_{\mathcal{I}_{\bar b, k}}\,\hat {\mathbf{t}}^{(Q)}_{\bar b, k,s}
    \in \mathbb{C}^{M \times 1}.
\end{align}
The reconstructed vectors $\{\tilde{\mathbf{f}}_{\bar b, k,s}\}$ capture the dominant subspace of the effective channels, enabling efficient precoding design with reduced feedback.

\subsection{Precoding Design at the BS}
In this section, we discuss the BS precoder design under two ECSI availability scenarios: (i) when the UE feeds back only the effective channel corresponding to its serving BS, and (ii) when the UE additionally feeds back the effective channels corresponding to interfering BSs. Let $\mathcal{B}_k$ denote the set of BSs whose ECSI is available for UE~$k$. If $\mathcal{B}_k = \{b_k\}$, equivalently $|\mathcal{B}_k| = 1$, only serving-cell ECSI is available and inter-cell interference is neglected. If $|\mathcal{B}_k| > 1$, ECSI from a subset of non-serving BSs is also available with additional feedback from the UE, enabling inter-cell interference-aware precoding. From the UE feedback, the BS reconstructs the per-stream effective channel $\tilde{\mathbf f}_{\bar b,k,s}$ as in~\eqref{eq:rec} and computes the precoder for stream~$s$ of UE~$k$ as
\begin{align}
\label{eq:bsbf_effCSI_int}
\mathbf w_{k,s}
=
\Bigg(
\sum_{\bar b \in \mathcal{B}_k}
\sum_{ j \in \mathcal K_{\bar b}}
\sum_{\bar s \in \mathcal S_j}
\tilde{\mathbf f}_{\bar b,j,\bar s}
\tilde{\mathbf f}_{\bar b,j,\bar s}^{\herm}
+ \nu_b \mathbf I_M
\Bigg)^{-1}
\tilde{\mathbf f}_{b_k,k,s},
\end{align}
where $\nu_b \ge 0$ is chosen to satisfy the BS transmit power constraint. Since the outer summation runs over $\mathcal{B}_k$, expression~\eqref{eq:bsbf_effCSI_int} covers both availability scenarios: inter-cell interference is explicitly suppressed when $|\mathcal{B}_k| > 1$, and only the intra-cell terms remain when $\mathcal{B}_k = \{b_k\}$.

After obtaining the initial precoders from Algorithm~\ref{alg:csi_int}, the BSs iteratively transmit UE-specific precoded pilots (e.g., DMRS), from which the UEs update their receive combiners. The updated combiners are then used to compute the ECSI, which is compressed, reported, and reconstructed for the subsequent precoder updates.  This iterative refinement forms an inner loop within the validity window of a given CSI-RS measurement, i.e., the CSI-RS-based channel estimate $\hat{\mathbf H}_{\bar b,k}$ is assumed to remain valid over the iterations, which holds as long as they span a fraction of the channel coherence time. The CSI-RS itself is updated in an outer loop, with a periodicity set by the channel dynamics, and the inner loop then continues from the updated estimate. The complete procedure for the refinement using the DMRS is summarized in Algorithm~\ref{alg:ecsi_int}, which can be implemented in either a distributed or a centralized manner. In a distributed implementation, the required ECSI information is exchanged among the cooperating BSs via backhaul links when $|\mathcal{B}_k|>1$. In a centralized implementation, the reconstructed ECSI is available at the central unit, which performs the precoder updates and distributes the resulting precoders to the corresponding BSs.

\begin{algorithm}[t]
\small
\begin{spacing}{1.2}
\textbf{Initialization:}
Obtain the initial precoders
$\{\mathbf w_{k,s}\}$
using Algorithm~\ref{alg:csi_int}.
\textbf{Repeat:}
\begin{enumerate}
\item Each BS transmits UE-specific precoded pilots
(e.g., DMRS)
as in~\eqref{eq:X_b_dl}, and each UE receives the pilots
from all BSs in $\mathcal B_k$
as in~\eqref{eq:Y_k_dl_2}.
\item Each UE updates its combiner
$\mathbf c_{k,s}$
using~\eqref{eq:rxmmse_est}.
\item Each UE computes the effective channels
$\hat{\mathbf f}_{\bar b,k,s}$,
compresses the ECSI into
$\hat{\mathbf t}^{(Q)}_{\bar b,k,s}$,
and reports the compressed ECSI together with the beam
indices $\mathcal I_{\bar b,k}$.
\item The effective channels
$\tilde{\mathbf f}_{\bar b,k,s}$
are reconstructed
as in~\eqref{eq:rec}
and made available for precoder design.
\item Update the precoders
$\{\mathbf w_{k,s}\}$
using~\eqref{eq:bsbf_effCSI_int}.
\end{enumerate}
\textbf{Until convergence}
\end{spacing}
\caption{Proposed OTA iterative beamforming design with ECSI feedback}
\label{alg:ecsi_int}
\end{algorithm}

\subsection{Effective Channel Error} \label{sec:eff_csi_error}
To compare conventional CSI and ECSI feedback under identical conditions, we quantify the error incurred in the \emph{post-combining effective channel} rather than the physical channel, since the effective channel determines the precoder design. The closed-form expressions for the effective channel error averaged jointly over channel realizations and UE combiners are intractable, as the combiner itself depends on multi-user interference, precoding, and channel statistics. We therefore condition on a fixed channel realization $\mathbf H_{\bar b,k}$ and a fixed UE combiner $\mathbf c_k$, and characterize the error averaged only over the receiver AWGN, which isolates the impact of beam selection, rank truncation, and channel-estimation noise while keeping the analysis tractable. Further, to ensure a consistent comparison between the two feedback schemes and to simplify the exposition, we restrict this AWGN-conditioned analysis to a single stream, omitting the stream index $s$ in what follows. Lastly, the sensitivity of the resulting errors to the specific choice of $\mathbf c_k$ is evaluated numerically in Section~\ref{sec:NUM}.

\subsubsection{Conventional CSI Feedback}
\label{subsec:raw_csi_error}
In this section, we analyze the effective channel error using the conventional CSI feedback discussed in Section~\ref{sec:CSI_comp}. Under conventional CSI feedback, the UE selects $L$ beams from noisy beam-domain channel estimates and feeds back a rank-$R$ compressed representation of the selected beam-domain channel. Let $\tilde{\mathbf V}_{\bar b,k}^{R} \in \mathbb{C}^{L\times R}$ and $\tilde{\mathbf U}_{\bar b,k}^{R} \in \mathbb{C}^{N\times R}$ denote the matrices containing the first $R$ columns of $\tilde{\mathbf V}_{\bar b,k}$ and $\tilde{\mathbf U}_{\bar b,k}$, respectively, and let $\tilde{\boldsymbol{\Lambda}}_{\bar b,k}^{R} \in \mathbb{C}^{R\times R}$ contain the first $R$ diagonal elements of $\tilde{\boldsymbol{\Lambda}}_{\bar b,k}$. Accordingly, the rank-$R$ approximation of the estimated beam-domain channel is $\tilde{\mathbf G}_{\bar b,k}^{R} \triangleq \tilde{\mathbf U}_{\bar b,k}^{R}\tilde{\boldsymbol{\Lambda}}_{\bar b,k}^{R}(\tilde{\mathbf V}_{\bar b,k}^{R})^{\herm}$.

To isolate the effects of beam selection, rank truncation, and AWGN-induced channel-estimation error, we neglect quantization error and assume perfectly known normalization factors, so that the reported $\bar{\mathbf V}_{\bar b,k}^{(Q)}$ at the BS can be replaced by its unquantized counterpart $\tilde{\mathbf V}_{\bar b,k}^{R}(\tilde{\boldsymbol{\Lambda}}_{\bar b,k}^{R})^{\herm}$. Since only right singular vectors are reported, the left singular vectors are never available at the BS. This implies that the BS has no information about the UE's actual combiner $\mathbf c_k$ or about $\tilde{\mathbf U}_{\bar b,k}^{R}$. However, to enable a fair comparison against the true-CSI case with the actual combiner, we form a rank-reduced virtual combiner from the true physical combiner $\mathbf c_k \in \mathbb{C}^{N\times 1}$ by projecting it onto the left-singular-vector basis as $(\tilde{\mathbf U}_{\bar b,k}^{R})^{\herm}\mathbf c_k$. This is the self-consistent choice available given the information retained by the rank-reduced representation. Applying this projected combiner to the rank-reduced CSI in \eqref{eq:com_rec_ch} gives the effective channel as
\begin{align}
\check{\mathbf f}_{\bar b,k}
&=\mathbf B_{\mathcal I_{\bar b,k}}\tilde{\mathbf V}_{\bar b,k}^{R}(\tilde{\boldsymbol{\Lambda}}_{\bar b,k}^{R})^{\herm}(\tilde{\mathbf U}_{\bar b,k}^{R})^{\herm}\mathbf c_k \nonumber \\
&=\mathbf B_{\mathcal I_{\bar b,k}}\left(\tilde{\mathbf G}_{\bar b,k}^{R}\right)^{\herm}\mathbf c_k,
\end{align}
while the true antenna-domain effective channel is $\mathbf f_{\bar b,k} = \mathbf H_{\bar b,k}^{\herm}\mathbf c_k$. Using the identity $\mathbf I=\mathbf B_{\mathcal I_{\bar b,k}}\mathbf B_{\mathcal I_{\bar b,k}}^{\herm}+\mathbf B_{\bar{\mathcal I}_{\bar b,k}}\mathbf B_{\bar{\mathcal I}_{\bar b,k}}^{\herm}$, with $\bar{\mathcal I}_{\bar b,k}\triangleq\{1,\dots,M\}\setminus\mathcal I_{\bar b,k}$, the error decomposes into three physically distinct contributions: a \emph{beam-selection} error $\mathbf e_{\bar b,k}^{\mathrm{beam}}$ from the $M-L$ discarded beams, a \emph{rank-truncation} error $\mathbf e_{\bar b,k}^{\mathrm{rank}}$ from the discarded singular modes of the $L$ selected beams, and an \emph{AWGN-induced} error $\mathbf e_{\bar b,k}^{\mathrm{awgn}}$ from noisy LS channel estimation perturbing the truncated SVD, given as
\begin{align}\label{eq:raw_error_decomp}
\mathbf e_{\bar b,k} &\triangleq \mathbf f_{\bar b,k}-\check{\mathbf f}_{\bar b,k} \nonumber \\
&=
\underbrace{\mathbf B_{\bar{\mathcal I}_{\bar b,k}}\mathbf B_{\bar{\mathcal I}_{\bar b,k}}^{\herm}\mathbf f_{\bar b,k}}_{\triangleq\, \mathbf e_{\bar b,k}^{\mathrm{beam}}}
+\underbrace{\mathbf B_{\mathcal I_{\bar b,k}}\left(\mathbf G_{\bar b,k}^{\bar R}\right)^{\herm}\mathbf c_k}_{\triangleq\, \mathbf e_{\bar b,k}^{\mathrm{rank}}} \nonumber \\
& \phantom{=}+\underbrace{\mathbf B_{\mathcal I_{\bar b,k}}\left(\mathbf G_{\bar b,k}^{R}-\tilde{\mathbf G}_{\bar b,k}^{R}\right)^{\herm}\mathbf c_k}_{\triangleq\, \mathbf e_{\bar b,k}^{\mathrm{awgn}}},
\end{align}
where $\mathbf G_{\bar b,k}^{R}$ is the noise-free rank-$R$ approximation of the beam-domain channel, and $\mathbf G_{\bar b,k}^{\bar R}\triangleq\mathbf G_{\bar b,k}-\mathbf G_{\bar b,k}^{R}$ is the residual discarded by rank truncation. Since $\mathbf e_{\bar b,k}^{\mathrm{beam}}$ lies in the subspace spanned by $\mathbf B_{\bar{\mathcal I}_{\bar b,k}}$, while $\mathbf e_{\bar b,k}^{\mathrm{rank}}$ and $\mathbf e_{\bar b,k}^{\mathrm{awgn}}$ lie in the orthogonal subspace spanned by $\mathbf B_{\mathcal I_{\bar b,k}}$, the beam-selection error is orthogonal to the other two, as a result
\begin{align}
    \mathbb E[\|\mathbf e_{\bar b,k}\|^2] = \mathbb E[\|\mathbf e_{\bar b,k}^{\mathrm{beam}}\|^2] + \mathbb E[\|\mathbf e_{\bar b,k}^{\mathrm{rank}}+\mathbf e_{\bar b,k}^{\mathrm{awgn}}\|^2].
\end{align}

We first evaluate the beam-selection error, due to the limited selection of beams based on noisy beam-domain energy estimate. From the LS estimate in~\eqref{eq:bd_ch}, the
$m$-th beam-domain channel estimate can be written~as
\begin{align}
\hat{\mathbf g}_{\bar b,k,m}
=
\mathbf g_{\bar b,k,m}
+
\frac{1}{\tau^{\beam}}
\mathbf Z_{k}^{\beam}
\tilde{\mathbf p}_m .
\end{align}
Assuming orthogonal pilots with unit-modulus symbols,
$\mathbf Z_{k}^{\beam}\tilde{\mathbf p}_m/\tau^{\beam}  \sim \mathcal{CN}(0, \frac{\sigma_{\ue}^2}{\tau^{\beam}}\mathbf I_N)$.
Based on the estimated beam-domain channel, the UE computes the beam-energy
metric $\hat{\alpha}_{\bar b,k,m}=\|\hat{\mathbf g}_{\bar b,k,m}\|^2,$
and selects the $L$ beams with the largest values of
$\hat{\alpha}_{\bar b,k,m}$, as discussed in
Section~\ref{sec:CSI_comp}. This random TOP-$L$ rule leads to an intractable
order-statistics characterization. For analytical tractability, we replace the
random TOP-$L$ cutoff by a deterministic threshold $T^\star$ such that the
\emph{expected} number of selected beams equals $L$, i.e.,
\begin{align}\label{eq:ave_prob}
\sum_{m=1}^{M}
\Pr\!\left(
\hat{\alpha}_{\bar b,k,m}\ge T^\star
\right)
=
L.
\end{align}
To evaluate the selection probability, we characterize the distribution of
$\hat{\alpha}_{\bar b,k,m}$. Since the randomness in
$\hat{\alpha}_{\bar b,k,m}$ is induced solely by the Gaussian noise
$\mathbf Z_{k}^{\beam}\tilde{\mathbf p}_m/\tau^{\beam}$, it follows that,
conditioned on the true beam-domain channel
$\mathbf g_{\bar b,k,m}$, the random variable
$\tau^{\beam}\hat{\alpha}_{\bar b,k,m}/\sigma_{\ue}^2$
follows a noncentral chi-square distribution with $2N$ degrees of freedom and
noncentrality parameter
$\gamma_m=\tau^{\beam}\alpha_{\bar b,k,m}/\sigma_{\ue}^2$, where
$\alpha_{\bar b,k,m}=\|\mathbf g_{\bar b,k,m}\|^2$~\cite{Her11}.
Accordingly, the marginal selection probability of beam $m$ is
\begin{align}\label{eq:pr_bs}
p_m
=
\Pr\!\left(
\hat{\alpha}_{\bar b,k,m}\ge T^\star
\right)
=
Q_N\!\left(
\sqrt{\gamma_m},
\sqrt{\frac{\tau^{\beam}T^\star}{\sigma_{\ue}^2}}
\right),
\end{align}
where $Q_N(\cdot,\cdot)$ denotes the generalized Marcum-$Q$ function of order
$N$. The threshold $T^\star$ is obtained using a bisection search satisfying
\eqref{eq:ave_prob}. Based on the selection probabilities, the average beam-selection projector is
\begin{align}
\mathcal P
\triangleq
\mathbb E\!\left[
\mathbf B_{\mathcal I_{\bar b,k}}
\mathbf B_{\mathcal I_{\bar b,k}}^{\herm}
\right]
=
\sum_{m=1}^{M}
p_m\,
\mathbf b_m
\mathbf b_m^{\herm},
\end{align}
which yields the mean-squared beam-selection error
\begin{align}
\mathbb E\!\left[
\|\mathbf e_{\bar b,k}^{\mathrm{beam}}\|^2
\right]
=
\mathbf f_{\bar b,k}^{\herm}
\bar{\mathcal P}
\mathbf f_{\bar b,k},
\end{align}
where
$\bar{\mathcal P}\triangleq\mathbf I-\mathcal P$.

We next evaluate the remaining two error components,
$\mathbf e_{\bar b,k}^{\mathrm{rank}}$
and
$\mathbf e_{\bar b,k}^{\mathrm{awgn}}$.
Both depend on the beam set selected in each noise realization.
To avoid enumerating all possible beam subsets together with their selection
probabilities, we evaluate these terms using a representative beam set
$\mathcal I_{\bar b,k}^{\star}$, chosen as the $L$ beams with the largest
selection probabilities. Moreover, AWGN perturbs not only the beam selection but also the singular subspace used for rank reduction.
Assuming sufficiently high pilot SNR, we adopt a first-order
subspace-perturbation approximation for
$\mathbf e_{\bar b,k}^{\mathrm{awgn}}$.
Under this approximation, the perturbation becomes a linear function of the LS
noise
$\mathbf Z_{k}^{\beam}
\tilde{\mathbf P}_{\mathcal I_{\bar b,k}^{\star}}/\tau^{\beam}$,
and is therefore uncorrelated with
$\mathbf e_{\bar b,k}^{\mathrm{rank}}$.
Consequently,
\begin{align}
\mathbb E\!\left[
\left\|
\mathbf e_{\bar b,k}^{\mathrm{rank}}
+
\mathbf e_{\bar b,k}^{\mathrm{awgn}}
\right\|^2
\right]
\approx
\mathbb E\!\left[
\|\mathbf e_{\bar b,k}^{\mathrm{rank}}\|^2
\right]
+
\mathbb E\!\left[
\|\mathbf e_{\bar b,k}^{\mathrm{awgn}}\|^2
\right].
\end{align}
For the representative beam set
$\mathcal I_{\bar b,k}^{\star}$,
the mean-squared rank-truncation error follows from
\eqref{eq:raw_error_decomp} as
\begin{align}
\mathbb E\!\left[
\|\mathbf e_{\bar b,k}^{\mathrm{rank}}\|^2
\right]
&=
\mathbb E\!\left[
\|
\mathbf c_k^{\herm}
\mathbf G_{\bar b,k}^{\bar R}
\|^2
\right] \\
&\approx
\sum_{r>R}
\lambda_r^2(\mathcal I_{\bar b,k}^{\star})
\left|
\mathbf u_r(\mathcal I_{\bar b,k}^{\star})^{\herm}
\mathbf c_k
\right|^2,
\end{align}
where
$\lambda_r(\mathcal I_{\bar b,k}^{\star})$
and
$\mathbf u_r(\mathcal I_{\bar b,k}^{\star})$
denote the singular values and corresponding left singular vectors of the
ideal beam-domain channel
$\mathbf G_{\bar b,k}$.
The rank-truncation error therefore depends not only on the discarded singular
values but also on the alignment between the combiner and the corresponding
left singular vectors.

We next evaluate the mean-squared AWGN-induced error.
From~\eqref{eq:bm_ch}, define the reduced LS estimation noise as
\begin{align}
\tilde{\mathbf Z}_{k,\mathcal I^\star_{\bar b,k}}
=
\frac{1}{\tau^{\beam}}
\mathbf Z_{k}^{\beam}
\tilde{\mathbf P}_{\mathcal I^\star_{\bar b,k}}
\in\mathbb C^{N\times L}.
\end{align}
At sufficiently high pilot SNR, the perturbation of the dominant singular
subspace is small. Using the first-order perturbation expansion of the truncated
SVD~\cite[Eq.~(13)]{Tru21}, and neglecting the higher-order cross terms
associated with singular-subspace rotations, the perturbation of the
rank-$R$ approximation can be expressed as
\begin{align}\label{eq:def_Ztilde_I_awgn}
\mathbf G_{\bar b,k}^{R}
-
\tilde{\mathbf G}_{\bar b,k}^{R}
\approx
\tilde{\mathbf Z}_{k,\mathcal I^\star_{\bar b,k}}
-
\mathbf T_{\mathbf U}^{\herm}
\tilde{\mathbf Z}_{k,\mathcal I^\star_{\bar b,k}}
\mathbf T_{\mathbf V},
\end{align}
where $\mathbf T_{\mathbf U}
\triangleq
\mathbf U_{\bar b,k}^{\bar R}
(\mathbf U_{\bar b,k}^{\bar R})^{\herm},
\qquad
\mathbf T_{\mathbf V}
\triangleq
\mathbf V_{\bar b,k}^{\bar R}
(\mathbf V_{\bar b,k}^{\bar R})^{\herm},$
with
$\mathbf U_{\bar b,k}^{\bar R}\in\mathbb C^{N\times(N-R)}$
and
$\mathbf V_{\bar b,k}^{\bar R}\in\mathbb C^{L\times(L-R)}$
containing the left and right singular vectors associated with the discarded
singular modes, respectively. Consequently, substituting~\eqref{eq:def_Ztilde_I_awgn} into the AWGN error component in
\eqref{eq:raw_error_decomp}, and using the
orthonormality of
$\mathbf B_{\mathcal I^\star_{\bar b,k}}$,
the mean-squared AWGN-induced error is obtained as
\begin{align}\label{eq:raw_awgn_reduced}
\mathbb E\!\left[
\|\mathbf e_{\bar b,k}^{\mathrm{awgn}}\|^2
\right]
=
\frac{\sigma_{\ue}^2}{\tau^{\beam}}
\mathbf c_k^{\herm}
\left(
L\mathbf I
-
(L-R)\mathbf T_{\mathbf U}
\right)
\mathbf c_k.
\end{align}

Combining the beam-selection, rank-truncation, and AWGN-induced components derived above yields the following central result.

\begin{proposition}\label{prop:mse_conventional}
The mean-squared effective-channel error under conventional CSI feedback is approximated as
\begin{align}\label{eq:csi_error}
\mathbb E\!\left[
\|\mathbf e_{\bar b,k}\|^2
\right]
\approx \
&
\mathbf f_{\bar b,k}^{\herm}
\bar{\mathcal P}
\mathbf f_{\bar b,k}
+
\sum_{r>R}
\lambda_r^2(\mathcal I^\star_{\bar b,k})
\left|
\mathbf u_r^{\herm}
\mathbf c_k
\right|^2
\nonumber\\
&
+
\frac{\sigma_{\ue}^2}{\tau^{\beam}}
\mathbf c_k^{\herm}
\left(
L\mathbf I
-
(L-R)\mathbf T_{\mathbf U}
\right)
\mathbf c_k.
\end{align}
\end{proposition}

The three terms in \eqref{eq:csi_error} correspond directly to beam selection, rank truncation, and AWGN, respectively. Notably, the rank-truncation term depends not only on the discarded singular values but also on how well the fixed combiner $\mathbf c_k$ aligns with the corresponding discarded left singular vectors.

\subsubsection{ECSI Feedback}
\label{subsec:ecsi_error_avg}
In this section, we analyze the effective-channel error under the proposed ECSI
feedback scheme, conditioning on a fixed channel realization $\H_{\bar b,k}$
and a fixed UE combiner $\c_{k}$.
The UE compresses the effective channel in the beam domain as in \eqref{eq:beam_eff}, using the beam index set
$\mathcal I_{\bar b,k}$ selected from noisy measurements, and the BS reconstructs
$\tilde{\mathbf f}_{\bar b,k}$ as in~\eqref{eq:rec}. Neglecting quantization error,
the reconstructed effective channel at the BS can be written as
\begin{align}\label{eq:ecsi_eff_est_decomp}
\tilde{\mathbf f}_{\bar b,k} =
   \mathbf{B}_{\mathcal{I}_{\bar b, k}}  \mathbf{B}_{\mathcal{I}_{\bar b, k}}^{\herm} 
   \underbrace{\H_{\bar b, k}^{\herm}\mathbf{c}_{k}}_{\triangleq\,\mathbf f_{\bar b,k}}
    + \mathbf{B}_{\mathcal{I}_{\bar b, k}} 
    \underbrace{\frac{1}{\tau^{\beam}}\,  
    \P_{\mathcal{I}_{\bar b, k}}^{\herm} (\Z_{k}^{\beam})^{\herm}  \mathbf{c}_{k}}_{\triangleq\,\mathbf n_{\bar b,k}}
\end{align}
where $\mathbf n_{\bar b,k}\in\mathbb C^{L\times 1}$ denotes the beam-domain
effective-channel estimation noise induced by $\Z_{k}^{\beam}$. Under orthogonal pilots,
$\mathbf n_{\bar b,k} \sim \mathcal{CN}(0, \sigma_{\ue}^2 \|\c_{k}\|^2 /\tau^{\beam} \, \I_L)$. Using the orthogonal beam decomposition as in the conventional CSI analysis,
the effective-channel reconstruction error becomes
\begin{align}
\check{\mathbf e}_{\bar b,k}
&\triangleq
\mathbf f_{\bar b,k} - \tilde{\mathbf f}_{\bar b,k} \nonumber\\
&=
\underbrace{\B_{\bar{\mathcal I}_{\bar b,k}}\B_{\bar{\mathcal I}_{\bar b,k}}^{\herm}
\mathbf f_{\bar b,k}}_{\triangleq\ \check{\mathbf e}_{\bar b,k}^{\mathrm{beam}}}
-
\underbrace{\B_{\mathcal I_{\bar b,k}}
\mathbf n_{\bar b,k}}_{\triangleq\, \check{\mathbf e}_{\bar b,k}^{\mathrm{awgn}}} .
\end{align}
Conditioned on $\mathcal I_{\bar b,k}$, 
$\check{\mathbf e}_{\bar b,k}^{\mathrm{awgn}}$ lies in the subspace spanned by 
$\B_{\mathcal I_{\bar b,k}}$, while 
$\check{\mathbf e}_{\bar b,k}^{\mathrm{beam}}$ lies in its orthogonal complement. 
Therefore, the two components are orthogonal. Consequently, the mean-squared error of the effective channel can be written as
\begin{align}\label{eq:ecsi_mse_decomp}
\Exp\!\left[\|\check{\mathbf e}_{\bar b,k}\|^2\right]
=
\Exp\!\left[\|\check{\mathbf e}_{\bar b,k}^{\mathrm{beam}} \|^2\right]
+
\Exp\!\left[\|\check{\mathbf e}_{\bar b,k}^{\mathrm{awgn}}\|^2\right].
\end{align}

We begin by evaluating the error component due to beam selection. The beam set $\mathcal I_{\bar b,k}$ is selected using the energy metric $\hat{\beta}_{\bar b,k,m}
\triangleq | \hat{\mathbf g}_{\bar b,k,m}^{\herm}\c_k |^{2},$
where the UE selects the $L$ beams with the largest $\hat{\beta}_{\bar b,k,m}$.
As in the conventional CSI analysis, to avoid intractable order statistics,
we replace the random TOP-$L$ rule by a deterministic threshold $\check{T}^\star$
such that the expected number of selected beams equals $L$, i.e.,
\begin{align}\label{eq:bs_ecsi}
\sum_{m=1}^M
\Pr\!\left(
\hat{\beta}_{\bar b,k,m}\ge \check{T}^\star
\right)
= L.
\end{align}
Since the randomness in $\hat{\beta}_{\bar b,k,m}$ is induced by the Gaussian
noise $\mathbf n_{\bar b,k}$, and conditioned on the true energy metric $\beta_{\bar b,k,m}
\triangleq \big| \mathbf g_{\bar b,k,m}^{\herm}\c_k \big|^{2},$ the random variable
$\hat{\beta}_{\bar b,k,m}/\sigma_{\beta}^2$, with
$\sigma_{\beta}^2
\triangleq
\sigma_{\ue}^2 \|\c_{k}\|^2/\tau^{\beam},$
follows a noncentral chi-square distribution with $2$ degrees of freedom and
noncentrality parameter $\check{\gamma}_m
\triangleq
{\beta}_{\bar b,k,m}/\sigma_{\beta}^2$ \cite{Her11}. Hence, the marginal selection probability of beam~$m$ is
\begin{align}\label{eq:ecsi_pr_sel}
\check{p}_m =
\Pr\!\left(
\hat{\beta}_{\bar b,k,m}\ge \check{T}^\star
\right)
=
Q_{1}\!\left(
\sqrt{\check{\gamma}_m},
\sqrt{ \frac{\check{T}^\star}{\sigma_{\beta}^2}}
\right).
\end{align}
The threshold $\check{T}^\star$ can be obtained via a bisection search satisfying
\eqref{eq:bs_ecsi}. Based on the selection probabilities, the average beam-selection projector is
\begin{align}
\check{\mathcal P}
\triangleq
\Exp\!\left[\B_{\mathcal I_{\bar b,k}}\B_{\mathcal I_{\bar b,k}}^{\herm}\right]
=
\sum_{m=1}^M \check{p}_m\,\b_m\b_m^{\herm},
\end{align}
which yields the mean-squared beam-selection error
\begin{align}
\Exp\!\left[\|\check{\mathbf e}_{\bar b,k}^{\mathrm{beam}}\|^2\right]
=
\mathbf f_{\bar b,k}^{\herm}
\bar{\check{\mathcal P}}
\,\mathbf f_{\bar b,k},
\end{align}
where $\bar{\check{\mathcal P}}\triangleq\I-\check{\mathcal P}$.

Next, we analyze the mean-squared AWGN-induced error as a function of the beam-selection probability. Since the beam set $\mathcal I_{\bar b,k}$ is selected based on the noisy beam-selection metrics $\hat{\beta}_{\bar b,k,m}$, the selected beams depend on the noise realization. Therefore, the AWGN-induced error must be evaluated conditioned on the beam-selection event. This conditioning is necessary because selecting beams with large noisy metrics introduces a statistical bias in the corresponding reported beam-domain coefficients, i.e., $\hat{\mathbf{t}}_{\bar b,k}$ in~\eqref{eq:beam_eff}. The noisy beam-domain coefficient corresponding to beam $m$ is
\begin{align}
\hat{\mathbf g}_{\bar b,k,m}^{\herm}\c_k
=
\mathbf g_{\bar b,k,m}^{\herm}\c_k
+ \tilde n_{\bar b,k,m},
\end{align}
where the scalar $\tilde n_{\bar b,k,m}$ is the noise corresponds to beam $m$, which is circularly symmetric complex
Gaussian with zero mean and variance $\sigma_{\beta}^2$. For a selected beam $m$ (i.e., $\hat{\beta}_{\bar b,k,m}\ge \check{T}^\star$),
the conditional noise energy in the reported beam-domain coefficient is
\begin{align}\label{eq:cond_mom}
\Exp\!\left[
|\tilde n_{\bar b,k,m}|^2
\ \Big|\ 
\hat{\beta}_{\bar b,k,m}\ge \check{T}^\star
\right].
\end{align}
Since $\hat{\beta}_{\bar b,k,m}$ follows a noncentral chi-square distribution
with $2$ degrees of freedom and noncentrality $\check{\gamma}_m$, the conditional
moment in~\eqref{eq:cond_mom} can be expressed in closed form using Nuttall-$Q$
functions as given in~\eqref{eq:ecsi_cond_noise_energy} \cite{Sun10}.
\begin{figure*}
\begin{align}\label{eq:ecsi_cond_noise_energy}
\Exp\!\left[
|\tilde n_{\bar b,k,m}|^2
\ \Big|\ 
\hat{\beta}_{\bar b,k,m}\ge \check{T}^\star
\right]
&=
\frac{\sigma_{\beta}^2}{Q_{1,0}(a_m,b)}
\Big(
Q_{3,0}(a_m,b)
-2a_m\,Q_{2,1}(a_m,b)
+a_m^2\,Q_{1,0}(a_m,b)
\Big)
\end{align}
\hrule
\end{figure*}
Here $a_m\triangleq \sqrt{2\check{\gamma}_m}$, $b \triangleq \sqrt{2{\check{T}^\star}/ \sigma_{\beta}^2}$,
and $Q_{i,j}(c,d)$ is the Nuttall-$Q$ function, defined as
\begin{align}
Q_{i,j}(c,d)
\triangleq
\int_{d}^{\infty}
x^{i}\exp\!\left(-\frac{x^{2}+c^{2}}{2}\right)
I_{j}(cx)\,dx,
\end{align}
with $I_j(\cdot)$ denoting the modified Bessel function of the first kind. Consequently, using the selection probabilities in~\eqref{eq:ecsi_pr_sel}, the
expected AWGN-induced error can be written as
\begin{align}\label{eq:ecsi_awgn_term}
\Exp\!\left[\|\check{\mathbf e}_{\bar b,k}^{\mathrm{awgn}}\|^2\right] \!
\approx \!
\sum_{m=1}^{M}
\check{p}_m\,
\Exp\!\left[
|\tilde n_{\bar b,k,m}|^2 \!
\ \Big|\ 
\hat{\beta}_{\bar b,k,m}\ge \check{T}^\star
\right].
\end{align}
For comparison with the conventional CSI analysis, we also consider the
high pilot-SNR regime. In this case, beam selection becomes effectively
deterministic for the dominant beams, and the conditioning effect vanishes.
Thus,
\begin{align}
\Exp\!\left[
|\tilde n_{\bar b,k,m}|^2
\ \Big|\ 
\hat{\beta}_{\bar b,k,m}\ge \check{T}^\star
\right]
\approx
\Exp\!\left[
|\tilde n_{\bar b,k,m}|^2
\right]
=
\sigma_{\beta}^2.
\end{align}

Combining the beam-selection and AWGN-induced components derived above yields the following central result.

\begin{proposition}\label{prop:mse_ecsi}
At sufficiently high pilot SNR, the mean-squared effective-channel error under ECSI feedback is approximated as
\begin{align}\label{eq:ecs_error}
\Exp\!\left[\| \check{\mathbf e}_{\bar b,k}\|^2\right]
\approx
\mathbf f_{\bar b,k}^{\herm}\bar{\check{\mathcal P}}\,
\mathbf f_{\bar b,k}
+
L \frac{\sigma_{\ue}^2}{\tau^{\beam}}\|\c_{k}\|^2.
\end{align}
\end{proposition}

The two terms in \eqref{eq:ecs_error} correspond to beam selection and AWGN, respectively; unlike conventional CSI feedback, no rank-truncation term appears, since ECSI compresses the post-combining effective channel directly without an explicit rank-reduction step.

Propositions~\ref{prop:mse_conventional} and~\ref{prop:mse_ecsi} give the effective-channel errors under conventional CSI and ECSI feedback, respectively. In both cases, the beam-selection error is governed by the average beam-selection projector and reflects the loss incurred by discarding the non-selected beam subspace. In conventional CSI feedback, the AWGN contribution is partially mitigated because the noise is effectively projected onto the retained low-dimensional subspace after rank truncation. This projection can reduce its impact compared to the full-dimensional noise component present in the ECSI scheme. However, conventional CSI feedback introduces an additional error term due to rank truncation arising from the SVD-based compression of the beam-domain channel. Although this rank-truncation error is typically small when the channel exhibits limited rank or strong beam sparsity, it critically depends on the alignment between the UE combiner and the omitted singular subspace. In particular, if the combiner has a significant projection onto the discarded subspace, the resulting error may become non-negligible. Such a situation can occur in interference-limited scenarios, where the UE combiner is shaped primarily by interference suppression rather than alignment with the dominant channel modes. In contrast, the proposed ECSI feedback avoids explicit rank truncation by directly compressing the post-combining effective channel. As a result, its error structure consists solely of beam-selection and AWGN-induced components. In the following, we discuss practical implementation aspects of the proposed framework and analyze the associated feedback overhead.

\begin{remark}
\textit{The proposed ECSI-based feedback framework naturally extends to multi-TRP coordinated transmission scenarios. In such settings, the UE can compute and report the effective CSI corresponding to multiple TRPs, i.e., for all $\bar b \in \mathcal{B}_{k}$, by applying the same post-combining operation. Since the ECSI represents the channel after UE combining, it provides a directly usable representation for joint precoder design across multiple TRPs without requiring any additional alignment. In contrast, extending conventional CSI feedback to multi-TRP scenarios is more challenging. In CSI-based methods, each BS–UE channel is compressed via a low-rank approximation obtained from a local SVD, which is defined up to a (local) unitary transformation. As a result, the corresponding dominant subspaces across different TRPs may not be aligned. Since there now exists unitary ambiguity, which is not explicitly fed back, combining CSI from multiple TRPs can lead to mismatches in the reconstructed channel representations, thereby degrading the performance of coordinated precoding. This limitation is avoided in the proposed ECSI framework, which directly operates on the effective channel observed after UE combining.}
\end{remark}

\subsection{Implementation and Feedback Overhead}
The proposed ECSI framework integrates naturally with existing 5G NR signaling procedures. The initialization phase relies on conventional CSI acquisition, obtained either via uplink pilots (e.g., SRS) or through codebook-based CSI-RS feedback, which the BS uses to design the initial precoders. Subsequent refinement of the precoders and combiners using the proposed ECSI method is performed through standard downlink pilots, such as DMRS, which are already multiplexed with data symbols. This enables iterative ECSI updates without incurring additional downlink signaling overhead. However, the ECSI needs to be reported using dedicated resources in the uplink, even if the channel is static, since each report reflects the most recent precoder--combiner pair. If the channel changes slowly, CSI-RS can be transmitted periodically in the downlink, allowing the UE to acquire the latest channel and provide the corresponding ECSI feedback, based on which the BS can design the precoder, as in Section~\ref{subsec:ecs_comp}. However, if channel conditions change significantly, CSI-RS transmissions can be used to reacquire updated CSI at both the BS and UE using conventional CSI feedback, as in Section~\ref{subsec:csi_comp}. Furthermore, the proposed feedback mechanism reuses the same payload structure as conventional type-II codebook-based reporting, thereby ensuring backward compatibility and straightforward integration with existing NR procedures.

To report the effective channel between BS~$\bar b$ and UE~$k$ at each iteration, the UE feeds back $L$ beam indices, where each beam index requires $\log_2 M$ bits. In addition, the feedback of $\hat{\mathbf{t}}_{\bar b, k,s}$ in~\eqref{eq:beam_eff} consists of $L \times 1$ complex effective-channel coefficients for each stream~$s$. The magnitude and phase of each coefficient are quantized using $\log_2 q_m$ and $\log_2 q_p$ bits, respectively, at each iteration. Furthermore, $S$ normalization coefficients are fed back, each requiring $\log_2 q_n$ bits. Therefore, the total feedback overhead of the proposed ECSI scheme is $L\log_2 M + S\log_2 q_n + LS(\log_2 q_m + \log_2 q_p)$ for each iteration, even if the channel is quasi-static, during the iterative update of the combiner and precoder when the channel changes slowly. In addition, if the channel changes significantly, conventional CSI feedback can be used instead, which reports a rank-$R$ representation, resulting in a total overhead of $L\log_2 M + R\log_2 q_n + LR(\log_2 q_m + \log_2 q_p)$.

\section{Numerical Results and Discussion} \label{sec:NUM}

In this section, we compare the performance of the proposed ECSI feedback-based precoding schemes presented in Section~\ref{sec:ECSI} (Algorithm~\ref{alg:ecsi_int}) with the CSI feedback-based precoding scheme described in Section~\ref{sec:CSI_comp} (Algorithm~\ref{alg:csi_int}), applicable to both TDD and frequency-FDD systems. Unless stated otherwise, the following parameters remain fixed across all simulations. The simulation setup consists of $B=4$ BSs, each equipped with $M=32$ antennas, placed on a square grid with an inter-site distance of $100$~m. A total of $K=8$ UEs, each with $N=6$ antennas, are uniformly distributed across the coverage area, and each UE is assigned $S=4$ data streams. The channel is modeled as $\H_{\bar b,k} = \delta_{\bar b,k}\tilde{\H}_{\bar b,k}$, where $\tilde{\H}_{\bar b,k}$ represents the small-scale fading generated using a one-ring model with a $20^\circ$ of angular spread, and multiple paths, a structure that can be efficiently represented by a small number of beams~\cite{Da00}. The large-scale fading coefficient is given by $\delta_{\bar b,k} = -48 - 30\log_{10}(d_{\bar b,k})$~[dB], where $d_{\bar b,k}$ is the distance between BS~$\bar b$ and UE~$k$. The BS transmit power is set to $\rho_b = 30$~dBm, while the UE AWGN variance is $\sigma^2_{\mathrm{UE}} = -95$~dBm. For feedback, each UE reports $L=8$ dominant beams and a maximum rank of $R=4$, with amplitude and phase quantized using $\log_2 q_m = \log_2 q_p = 4$ bits. The pilot lengths for cell-specific and UE-specific pilot transmissions are $\tau^{\beam} = 128$ and $\tau^{\pre} = 32$, respectively. As the performance metric, we evaluate the sum-rate in~\eqref{eq:sumrate}, averaged over $10^3$ independent channel realizations and UE placements.

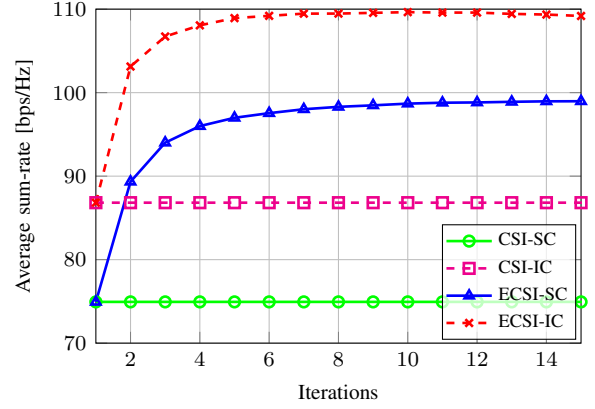
\begin{figure}[t!]
\begin{center}
\begin{tikzpicture}

\begin{axis}[
	width=8cm,
	height=6cm,
	xmin=1, xmax=15,
	ymin=70, ymax=110,
    xlabel={Iterations},
    ylabel={Average sum-rate [bps/Hz]},
    xlabel near ticks,
	ylabel near ticks,
    x label style={font=\footnotesize},
	y label style={font=\footnotesize},
    ticklabel style={font=\footnotesize},
   legend pos=south west,
    legend cell align=left,
    legend style={at={(0.99,0.01)}, anchor=south east},
	legend style={font=\scriptsize, inner sep=1pt, fill opacity=0.75, draw opacity=1, text opacity=1},
	grid=both,
]

\addplot[solid, line width=1pt, green, mark=o, mark options={solid}]
table[x=iter, y=cen,  , col sep=comma] 
{Figol/fig1data.txt};
\addlegendentry{{CSI-SC}};

\addplot[dashed, line width=1pt, magenta, mark=square, mark options={solid}]
table[x=iter, y=cen_int, col sep=comma] 
{Figol/fig1data.txt};
\addlegendentry{{CSI-IC}};

\addplot[solid, line width=1pt, blue,  mark=triangle, mark options={solid}]
table[x=iter, y=ecsi, col sep=comma] 
{Figol/fig1data.txt};
\addlegendentry{{ECSI-SC}};

\addplot[dashed, line width=1pt, red, mark=x, mark options={solid}]
table[x=iter, y=ecsi_int, col sep=comma] 
{Figol/fig1data.txt};
\addlegendentry{{ECSI-IC}};



\end{axis}

\end{tikzpicture}
\vspace{-2mm}
\caption{Average sum-rate over iterations.}
\label{fig:r_i}
\end{center}
\vspace{-4mm}
\end{figure}

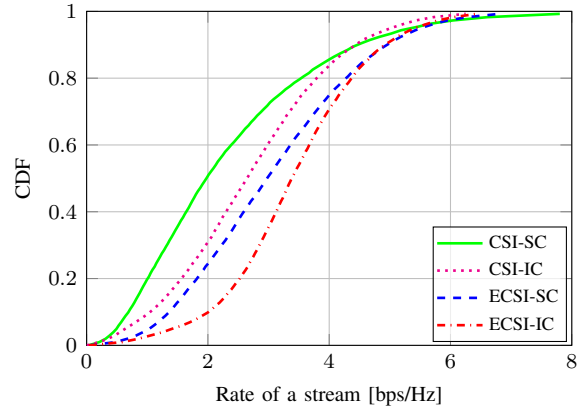
\begin{figure}[t!]
\begin{center}
\begin{tikzpicture}

\begin{axis}[
	width=8cm,
	height=6cm,
	xmin=0, xmax=8,
	ymin=0, ymax=1,
    ylabel={CDF},
    xlabel={Rate of a stream [bps/Hz]},
    xlabel near ticks,
	ylabel near ticks,
    x label style={font=\footnotesize},
	y label style={font=\footnotesize},
    ticklabel style={font=\footnotesize},
   legend pos=south west,
    legend cell align=left,
    legend style={at={(0.99,0.01)}, anchor=south east},
	legend style={font=\scriptsize, inner sep=1pt, fill opacity=0.75, draw opacity=1, text opacity=1},
	grid=both,
]

\addplot[solid, line width=1pt, green, mark options={solid}]
table[x=yCSI, y=xCSI,   col sep=comma] 
{Figol/fig2datacsi.txt};
\addlegendentry{{CSI-SC}};

\addplot[dotted, line width=1pt, magenta, mark options={solid}]
table[x=yCSIInt, y=xCSIInt, col sep=comma] 
{Figol/fig2datacsiInt.txt};
\addlegendentry{{CSI-IC}};

\addplot[dashed, line width=1pt, blue, mark options={solid}]
table[x=yECSI, y=xECSI, col sep=comma] 
{Figol/fig2dataecsi.txt};
\addlegendentry{{ECSI-SC}};

\addplot[dash dot, line width=1pt, red, mark options={solid}]
table[x=yECSIInt, y=xECSIInt, col sep=comma] 
{Figol/fig2dataecsiint.txt};
\addlegendentry{{ECSI-IC}};

\end{axis}

\end{tikzpicture}
\vspace{-2mm}
\caption{Rate distribution of UE streams.}
\label{fig:cdf_r}
\end{center}
\vspace{-4mm}
\end{figure}
In Fig.~\ref{fig:r_i}, we plot the sum rate of the proposed ECSI-based precoding schemes with and without inter-cell coordination, denoted as ECSI serving-cell (ECSI-SC) and ECSI inter-cell coordination (ECSI-IC), respectively, as a function of OTA iterations. These results are compared with the conventional precoding designs based on CSI feedback, denoted as CSI serving-cell (CSI-SC) and CSI inter-cell coordination (CSI-IC), respectively. The proposed ECSI-SC and ECSI-IC schemes achieve approximately $30\%$ and $25\%$ higher sum rate than their conventional counterparts CSI-SC and CSI-IC, respectively. Moreover, the ECSI-based methods converge rapidly, saturating within about five iterations. The coordinated schemes, CSI-IC and ECSI-IC, consistently outperform their non-coordinated counterparts by explicitly accounting for inter-cell interference and proactively suppressing interference toward UEs in neighboring cells. These gains are most pronounced for cell-edge UEs, as illustrated in Fig.~\ref{fig:cdf_r}. In particular, at the 10th percentile of the per-stream rate distribution (i.e., considering the worst $10\%$ of streams), CSI-IC and ECSI-IC achieve approximately $42\%$ and $48\%$ higher per stream rate, respectively, compared to CSI-SC and ECSI-SC. Similarly, ECSI-SC and ECSI-IC provide up to a $100\% $ improvement over their corresponding conventional CSI-based schemes. This performance improvement stems from the fact that conventional CSI feedback limits the available spatial degrees of freedom through channel compression, which restricts the ability of the precoder–combiner pair to fully mitigate inter-cell interference. In contrast, the proposed ECSI framework allows the UE combiner to exploit all available spatial degrees of freedom in the effective channel, thereby enabling more effective interference suppression and improved SINR.

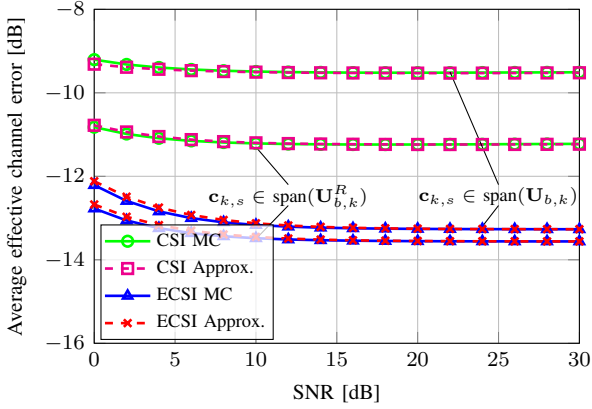
\begin{figure}[t!]
\begin{center}
\begin{tikzpicture}

\begin{axis}[
	width=8cm,
	height=6cm,
	xmin=0, xmax=30,
	ymin=-16, ymax=-8,
    xlabel={SNR [dB]},
    ylabel={Average effective channel error [dB]},
    xlabel near ticks,
	ylabel near ticks,
    x label style={font=\footnotesize},
	y label style={font=\footnotesize},
    ticklabel style={font=\footnotesize},
   legend pos=south west,
    legend cell align=left,
    legend style={at={(0.37,0.01)}, anchor=south east},
	legend style={font=\scriptsize, inner sep=1pt, fill opacity=0.75, draw opacity=1, text opacity=1},
	grid=both,
]

\addplot[solid, line width=1pt, green,  mark=o, mark options={solid}]
table[x=SNRdB, y=CSI_MC,  , col sep=comma] 
{Figol/fig3data.txt};
\addlegendentry{{CSI MC}};

\addplot[dashed, line width=1pt, magenta, mark=square, mark options={solid}]
table[x=SNRdB, y=CSI_TH, col sep=comma] 
{Figol/fig3data.txt};
\addlegendentry{{CSI Approx.}};

\addplot[solid, line width=1pt, blue, mark = triangle, mark options={solid}]
table[x=SNRdB, y=ECSI_MC, col sep=comma] 
{Figol/fig3data.txt};
\addlegendentry{{ECSI MC}};

\addplot[dashed, line width=1pt, red, mark= x, mark options={solid}]
table[x=SNRdB, y=ECSI_TH, col sep=comma] 
{Figol/fig3data.txt};
\addlegendentry{{ECSI Approx.}};

\addplot[solid, line width=1pt, green, mark=o,mark options={solid}]
table[x=SNRdB, y=CSI_MC,  , col sep=comma] 
{Figol/fig4data.txt};

\addplot[dashed, line width=1pt, magenta, mark=square, mark options={solid}]
table[x=SNRdB, y=CSI_TH, col sep=comma] 
{Figol/fig4data.txt};

\addplot[solid, line width=1pt, blue, mark=triangle, mark options={solid}]
table[x=SNRdB, y=ECSI_MC, col sep=comma] 
{Figol/fig4data.txt};

\addplot[dashed, line width=1pt, red, mark=x, mark options={solid}]
table[x=SNRdB, y=ECSI_TH, col sep=comma] 
{Figol/fig4data.txt};

\begin{scope}[>=latex]
\draw[-] (10,-11.3) -- (12,-12.2) {};
\draw[-] (10,-13.5) -- (12,-12.7) {};
\end{scope}

\node[black, font=\scriptsize] at (12,-12.5) {$\c_{k,s} \in $ span$( \U_{b, k}^{R})$};

\begin{scope}[>=latex]
\draw[-] (22,-9.5) -- (25,-12.2) {};
\draw[-] (24,-13.2) -- (25,-12.7) {};
\end{scope}

\node[black, font=\scriptsize] at (25,-12.5) {$\c_{k,s} \in $ span$(\U_{b, k})$};

\end{axis}

\end{tikzpicture}
\vspace{-2mm}
\caption{Effective channel error over SNR.}
\label{fig:ecsi_snr}
\end{center}
\vspace{-4mm}
\end{figure}

In Fig.~\ref{fig:ecsi_snr}, we compare the effective-channel estimation error obtained with CSI feedback and the proposed ECSI feedback as a function of the SNR. Both analytical approximations and Monte Carlo simulation results are presented, denoted as CSI Approx, CSI MC, ECSI Approx, and ECSI MC. For simplicity, we consider a single BS and a single UE, and average the results over $10^3$ independent channel realizations and the corresponding UE combiners. In this setting, firstly, we assume that the UE combiner lies within the subspace retained by the compressed CSI, i.e., $\c_{k,s} \in $ span$( \U_{b, k}^{R})$, and therefore the rank-truncation error in the CSI-SC method is zero (see \eqref{eq:csi_error}). Nevertheless, the beams selected based on conventional CSI are not necessarily optimal to represent the effective channel as in ECSI. As a result, the proposed ECSI feedback achieves an effective-channel estimation error that is approximately $2$~dB lower than that of CSI feedback across the considered SNR range. However, when the UE combiner spans the subspace of the true channel, i.e., $\c_{k,s} \in $ span$( \U_{b, k}^{R})$, the rank truncation in the CSI feedback induces a non-negligible error, which significantly increases the overall effective-channel estimation error. In contrast, the ECSI feedback method is robust to this effect, as it directly compresses the post-combining effective channel and does not rely on a low-rank approximation of the beam-domain channel. Consequently, a pronounced performance gap is observed between CSI and ECSI feedback, with ECSI achieving an error reduction of approximately $4$~dB. This scenario further highlights the practical relevance of ECSI feedback in interference-limited settings, where the UE combiner may deliberately span a larger portion of the channel subspace to suppress interference and enhance the desired signal, particularly for cell-edge users. These improvements in effective-channel estimation accuracy directly translate into more efficient precoder–combiner design and higher achievable downlink rates.

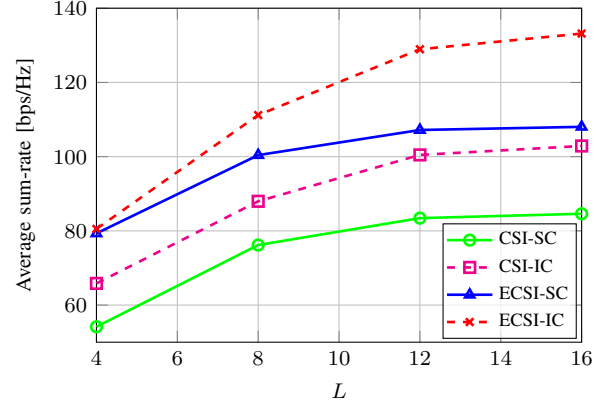
\begin{figure}[t!]
\begin{center}
\begin{tikzpicture}

\begin{axis}[
	width=8cm,
	height=6cm,
	xmin=4, xmax=16,
	ymin=50, ymax=140,
    xlabel={$L$},
    ylabel={Average sum-rate [bps/Hz]},
    xlabel near ticks,
	ylabel near ticks,
    x label style={font=\footnotesize},
	y label style={font=\footnotesize},
    ticklabel style={font=\footnotesize},
   legend pos=south west,
    legend cell align=left,
    legend style={at={(0.99,0.01)}, anchor=south east},
	legend style={font=\scriptsize, inner sep=1pt, fill opacity=0.75, draw opacity=1, text opacity=1},
	grid=both,
]

\addplot[solid, line width=1pt, green, mark=o, mark options={solid}]
table[x=beams, y=cen,  , col sep=comma] 
{Figol/fig5data.txt};
\addlegendentry{{CSI-SC}};

\addplot[dashed, line width=1pt, magenta, mark=square, mark options={solid}]
table[x=beams, y=cen_int, col sep=comma] 
{Figol/fig5data.txt};
\addlegendentry{{CSI-IC}};

\addplot[solid, line width=1pt, blue, mark=triangle, mark options={solid}]
table[x=beams, y=ecsi, col sep=comma] 
{Figol/fig5data.txt};
\addlegendentry{{ECSI-SC}};

\addplot[dashed, line width=1pt, red, mark=x, mark options={solid}]
table[x=beams, y=ecsi_int, col sep=comma] 
{Figol/fig5data.txt};
\addlegendentry{{ECSI-IC}};



\end{axis}

\end{tikzpicture}
\vspace{-2mm}
\caption{Sum-rate versus number of beams.}
\label{fig:r_vs_l}
\end{center}
\vspace{-4mm}
\end{figure}

In Fig.~\ref{fig:r_vs_l}, we illustrate the average sum-rate performance of all methods as a function of the number of reported beams $L$. As expected, the performance of all schemes improves as $L$ increases, since a larger beam set provides a more accurate representation of the serving BS channel in the CSI-SC and ECSI-SC schemes, as well as both the serving and interfering BS channels in the coordinated CSI-IC and ECSI-IC schemes. When only a small $L$ is reported, the dominant impairment is the self-interference caused by channel estimation and compression errors, which outweighs the impact of inter-cell interference. As a result, limited performance gain is observed from inter-cell coordination when only four beams are reported. As $L$ increases, the accuracy of both the desired and interfering channel representations improves. However, at large $L$, the performance of the non-coordinated CSI-SC and ECSI-SC schemes remains limited by residual inter-cell interference. In contrast, the coordinated schemes CSI-IC and ECSI-IC continue to benefit from increasing $L$, as improved serving and interference channel accuracy enables more effective interference suppression. Consequently, the performance gap between the coordinated and non-coordinated schemes widens as $L$ increases.

\begin{figure}[t!]
\begin{center}
\begin{tikzpicture}

\begin{axis}[
	width=8cm,
	height=6cm,
	xmin=8, xmax=64,
	ymin=30, ymax=120,
    xlabel={$M$},
    ylabel={Average sum-rate [bps/Hz]},
    xlabel near ticks,
	ylabel near ticks,
    x label style={font=\footnotesize},
	y label style={font=\footnotesize},
    ticklabel style={font=\footnotesize},
   legend pos=south west,
    legend cell align=left,
    legend style={at={(0.99,0.01)}, anchor=south east},
	legend style={font=\scriptsize, inner sep=1pt, fill opacity=0.75, draw opacity=1, text opacity=1},
	grid=both,
]

\addplot[solid, line width=1pt, green, mark=o, mark options={solid}]
table[x=antennas, y=cen,  , col sep=comma] 
{Figol/fig6data.txt};
\addlegendentry{{CSI-SC}};

\addplot[dashed, line width=1pt, magenta, mark=square, mark options={solid}]
table[x=antennas, y=cen_int, col sep=comma] 
{Figol/fig6data.txt};
\addlegendentry{{CSI-IC}};

\addplot[solid, line width=1pt, blue, mark=triangle, mark options={solid}]
table[x=antennas, y=ecsi, col sep=comma] 
{Figol/fig6data.txt};
\addlegendentry{{ECSI-SC}};

\addplot[dashed, line width=1pt, red, mark=x, mark options={solid}]
table[x=antennas, y=ecsi_int, col sep=comma] 
{Figol/fig6data.txt};
\addlegendentry{{ECSI-IC}};



\end{axis}

\end{tikzpicture}
\vspace{-2mm}
\caption{Sum-rate versus number of antennas at the BS}
\label{fig:r_vs_M}
\end{center}
\vspace{-4mm}
\end{figure}
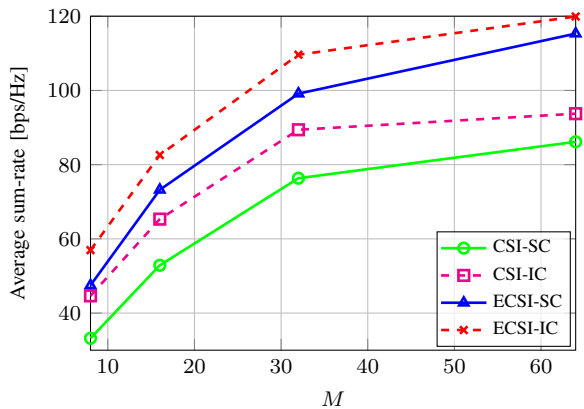

In Fig.~\ref{fig:r_vs_M}, we illustrate the average sum-rate performance of all methods as a function of the number of antennas $M$ at each BS. As expected, the performance of all schemes improves with increasing $M$ due to enhanced beamforming gain and improved spatial resolution. For a $M$, the relative gain achieved by inter-cell coordination becomes smaller compared to the low-antenna regime, since the channels associated with different cells become more spatially separable and inter-cell interference is naturally reduced.

\begin{figure}[t!]
\begin{center}
\begin{tikzpicture}

\begin{axis}[
	width=8cm,
	height=6cm,
	xmin=4, xmax=32,
	ymin=45, ymax=150,
    xlabel={$K$},
    ylabel={Average sum-rate [bps/Hz]},
    xlabel near ticks,
	ylabel near ticks,
    x label style={font=\footnotesize},
	y label style={font=\footnotesize},
    ticklabel style={font=\footnotesize},
   legend pos=south west,
    legend cell align=left,
    legend style={at={(0.99,0.01)}, anchor=south east},
	legend style={font=\scriptsize, inner sep=1pt, fill opacity=0.75, draw opacity=1, text opacity=1},
	grid=both,
]

\addplot[solid, line width=1pt, green, mark=o, mark options={solid}]
table[x=Ues, y=cen,  , col sep=comma] 
{Figol/fig7data.txt};
\addlegendentry{{CSI-SC}};

\addplot[dashed, line width=1pt, magenta, mark=square, mark options={solid}]
table[x=Ues, y=cen_int, col sep=comma] 
{Figol/fig7data.txt};
\addlegendentry{{CSI-IC}};

\addplot[solid, line width=1pt, blue, mark=triangle, mark options={solid}]
table[x=Ues, y=ecsi, col sep=comma] 
{Figol/fig7data.txt};
\addlegendentry{{ECSI-SC}};

\addplot[dashed, line width=1pt, red, mark=x, mark options={solid}]
table[x=Ues, y=ecsi_int, col sep=comma] 
{Figol/fig7data.txt};
\addlegendentry{{ECSI-IC}};



\end{axis}

\end{tikzpicture}
\vspace{-2mm}
\caption{Sum-rate versus total number of UEs.}
\label{fig:r_vs_k}
\end{center}
\vspace{-4mm}
\end{figure}
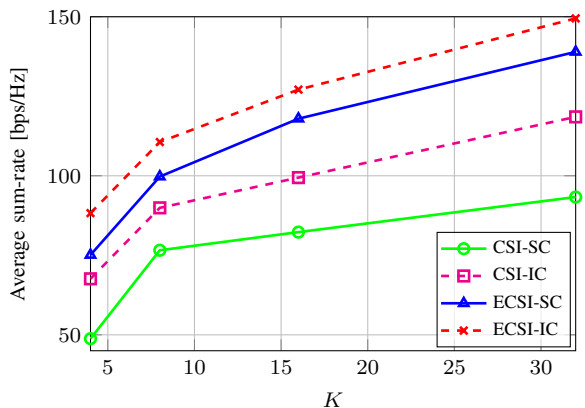

In Fig.~\ref{fig:r_vs_k}, we illustrate the average sum-rate performance of all methods as a function of the number of UEs $K$. As expected, the sum rate increases with $K$ due to the availability of additional spatial streams. However, a larger $K$ also intensifies inter-user and inter-cell interference. The CSI-SC, ECSI-SC schemes primarily mitigate inter-user interference, whereas the CSI-IC, ECSI-IC schemes address both inter-user and inter-cell interference. Under the proposed ECSI framework, the UE combiner is designed over the full subspace of the effective channel, enabling more efficient interference suppression and improved SINR. This added flexibility allows ECSI-based schemes to achieve larger performance gains compared to conventional CSI only feedback, particularly in dense user scenarios.

\begin{figure}[t!]
\begin{center}
\begin{tikzpicture}

\begin{axis}[
	width=8cm,
	height=6cm,
	xmin=2, xmax=5,
	ymin=60, ymax=120,
    xtick={2,3,4,5},
    xlabel={$q_m$ and $q_p$},
    ylabel={Average sum-rate [bps/Hz]},
    xlabel near ticks,
	ylabel near ticks,
    x label style={font=\footnotesize},
	y label style={font=\footnotesize},
    ticklabel style={font=\footnotesize},
   legend pos=south west,
    legend cell align=left,
    legend style={at={(0.28,0.64)}, anchor=south east},
	legend style={font=\scriptsize, inner sep=1pt, fill opacity=0.75, draw opacity=1, text opacity=1},
	grid=both,
]

\addplot[solid, line width=1pt, green, mark=o, mark options={solid}]
table[x=levels, y=cen,  , col sep=comma] 
{Figol/fig8data.txt};
\addlegendentry{{CSI-SC}};

\addplot[dashed, line width=1pt, magenta, mark=square, mark options={solid}]
table[x=levels, y=cen_int, col sep=comma] 
{Figol/fig8data.txt};
\addlegendentry{{CSI-IC}};

\addplot[solid, line width=1pt, blue, mark=triangle, mark options={solid}]
table[x=levels, y=ecsi, col sep=comma] 
{Figol/fig8data.txt};
\addlegendentry{{ECSI-SC}};

\addplot[dashed, line width=1pt, red, mark=x, mark options={solid}]
table[x=levels, y=ecsi_int, col sep=comma] 
{Figol/fig8data.txt};
\addlegendentry{{ECSI-IC}};



\end{axis}

\end{tikzpicture}
\vspace{-2mm}
\caption{Sum-rate versus quantization bits.}
\label{fig:r_vs_q}
\end{center}
\vspace{-4mm}
\end{figure}
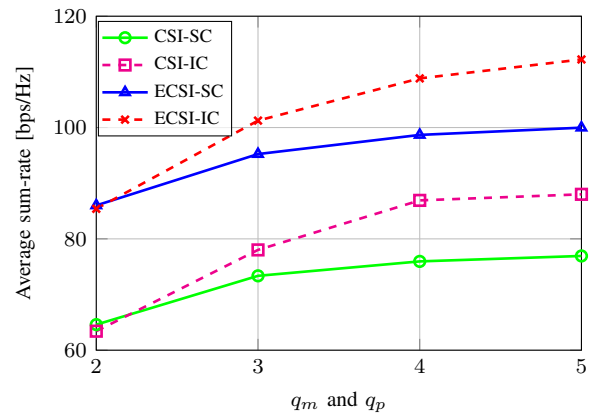

In Fig.~\ref{fig:r_vs_q}, we compare the average sum-rate performance of all methods as a function of the number of bits used for magnitude and phase quantization, $q_m$ and $q_p$, respectively. As expected, the sum rate of all schemes increases with the quantization resolution, since finer quantization reduces the feedback error associated with the reported beam-domain coefficients for a fixed beam set and rank truncation. When $q_m$ and $q_p$ is small, self-interference caused by quantization errors dominates over inter-cell interference, and therefore the coordinated schemes CSI-IC and ECSI-IC do not provide gains over their non-coordinated counterparts. As $q_m$ and $q_p$ increases, the quantization error is reduced, which simultaneously mitigates self-interference and enables the coordinated schemes to more effectively suppress inter-cell interference. Consequently, both CSI-IC and ECSI-IC achieve increasing performance gains over the non-coordinated CSI-SC and ECSI-SC schemes at higher quantization resolutions.

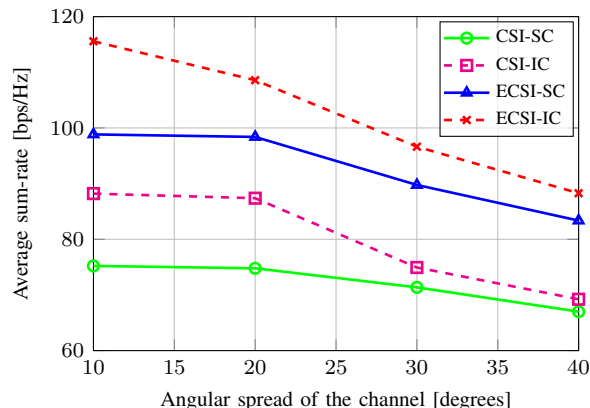
\begin{figure}[t!]
\begin{center}
\begin{tikzpicture}

\begin{axis}[
	width=8cm,
	height=6cm,
	xmin=10, xmax=40,
	ymin=60, ymax=120,
    xlabel={Angular spread of the channel [degrees]},
    ylabel={Average sum-rate [bps/Hz]},
    xlabel near ticks,
	ylabel near ticks,
    x label style={font=\footnotesize},
	y label style={font=\footnotesize},
    ticklabel style={font=\footnotesize},
   legend pos=south west,
    legend cell align=left,
    legend style={at={(0.99,0.64)}, anchor=south east},
	legend style={font=\scriptsize, inner sep=1pt, fill opacity=0.75, draw opacity=1, text opacity=1},
	grid=both,
]

\addplot[solid, line width=1pt, green, mark=o,mark options={solid}]
table[x=spread, y=cen,  , col sep=comma] 
{Figol/fig9data.txt};
\addlegendentry{{CSI-SC}};

\addplot[dashed, line width=1pt, magenta, mark=square, mark options={solid}]
table[x=spread, y=cen_int, col sep=comma] 
{Figol/fig9data.txt};
\addlegendentry{{CSI-IC}};

\addplot[solid, line width=1pt, blue, mark=triangle, mark options={solid}]
table[x=spread, y=ecsi, col sep=comma] 
{Figol/fig9data.txt};
\addlegendentry{{ECSI-SC}};

\addplot[dashed, line width=1pt, red, mark=x, mark options={solid}]
table[x=spread, y=ecsi_int, col sep=comma] 
{Figol/fig9data.txt};
\addlegendentry{{ECSI-IC}};



\end{axis}

\end{tikzpicture}
\vspace{-2mm}
\caption{Sum-rate versus channel spread.}
\label{fig:r_vs_spread}
\end{center}
\vspace{-4mm}
\end{figure}

In Fig.~\ref{fig:r_vs_spread}, we illustrate the average sum-rate performance of all methods as a function of the angular spread of the channel. A multipath channel model is considered, and therefore even at relatively small angular spreads the channel exhibits sufficient rank to support all spatial streams. As the angular spread increases, a larger number of angular components
is required to accurately represent the channel. The compression error
therefore grows with the angular spread for a fixed number of feedback
beams $L = 8$, and the CSI available at the transmitter becomes
progressively less accurate. Consequently, the precoders are less
effective at suppressing inter-user and inter-cell interference, an
effect further aggravated by the increasing overlap between the angular
supports of different UEs and cells, which reduces the achievable sum
rate for all schemes.

\section{Conclusion} \label{sec:CONC}
This paper proposed an ECSI feedback framework for multi-user MIMO downlink precoding in multi-cell coordinated systems. Unlike conventional codebook-based CSI feedback, where UEs report a compressed representation of the physical channel, the proposed approach enables UEs to feed back compact representations of their post-combining effective channels, thereby inherently accounting for the UE combining operation and eliminating rank-truncation mismatch. The proposed scheme reuses the existing codebook feedback structure and integrates naturally with OTA iterative refinement using standard downlink pilots, enabling improved alignment between BS precoders and UE combiners without introducing additional downlink signaling overhead. Analytical error characterizations highlight the fundamental differences between conventional CSI and ECSI feedback in terms of beam selection, rank truncation, and AWGN-induced errors. Simulation results demonstrate rapid convergence within a few iterations and show that ECSI achieves up to 30\% sum-rate improvement over conventional CSI-based precoding, with particularly strong gains for cell-edge and interference-limited UEs across a wide range of system and channel parameters.

\bibliographystyle{IEEEtran}
\bibliography{IEEEabbr,ref2}

@STRING{J_FNT_SP			= "Found. and Trends\textregistered~Signal Process."}

@STRING{J_IEEE_ACCESS		= "IEEE Access"}

@STRING{J_IEEE_TC      		= "IEEE Trans. Commun."}

@STRING{J_IEEE_CST			= "IEEE Commun. Surveys Tuts."}

@STRING{J_IEEE_TIT			= "IEEE Trans. Inf. Theory"}

@STRING{J_IEEE_JSAC			= "IEEE J. Sel. Areas Commun."}

@STRING{J_IEEE_TSP			= "IEEE Trans. Signal Process."}

@STRING{J_IEEE_WC			= "IEEE Wireless Commun."}

@STRING{J_IEEE_TWC			= "IEEE Trans. Wireless Commun."}

@STRING{M_IEEE_CM			= "IEEE Commun. Mag."}

@STRING{L_IEEE_CL			= "IEEE Commun. Lett."}

@STRING{L_IEEE_WCL			= "IEEE Wireless Commun. Lett."}

@STRING{C_IEEE_ICC			= "Proc. IEEE Int. Conf. Commun. (ICC)"}

@STRING{C_EUCNC_6GSUMMIT = "Proc. European Conf. Netw. Commun. (EuCNC) \& 6G Summit"}

@book{dahlman20book,
  author    = {E. Dahlman and S. Parkvall and J. Skold},
  title     = {{5G NR}: The Next Generation Wireless Access Technology},
  publisher = {Academic Press},
  year      = {2020}
}

@misc{Raj20,
  author       = {N. Rajatheva and I. Atzeni and E. Bj{\"o}rnson and others},
  title        = {White Paper on Broadband Connectivity in {6G}},
  year         = {2020},
  howpublished = {White paper}
}

@article{Mar10,
  author  = {T. L. Marzetta},
  title   = {Noncooperative cellular wireless with unlimited numbers of base station antennas},
  journal = J_IEEE_TWC,
  year    = {2010},
  volume  = {9},
  number  = {11},
  pages   = {3590--3600},
  month   = nov
}

@article{Bjornson17MassiveMIMOBook,
  author  = {E. Bj{\"o}rnson and J. Hoydis and L. Sanguinetti},
  title   = {Massive {MIMO} networks: Spectral, energy, and hardware efficiency},
  journal = J_FNT_SP,
  year    = {2017},
  volume  = {11},
  number  = {3--4},
  pages   = {154--655}
}

@TechReport{TS38211,
  author      = "{3GPP}",
  title       = {{NR}; Physical Channels and Modulation ({Release}~18)},
  institution = "{3GPP}",
  year        = {2025},
  number      = {TS 38.211 V18.7.0 (2025-07)}
}

@article{Shi11WMMSE,
  author  = {Q. Shi and M. Razaviyayn and Z.-Q. Luo and C. He},
  title   = {An iteratively weighted {MMSE} approach to distributed sum-utility maximization for a {MIMO} interfering broadcast channel},
  journal = J_IEEE_TSP,
  year    = {2011},
  volume  = {59},
  number  = {9},
  pages   = {4331--4340},
  month   = sep
}

@article{kom13,
  author  = {P. Komulainen and A. T{\"o}lli and M. Juntti},
  title   = {Effective {CSI} signaling and decentralized beam coordination in {TDD} multi-cell {MIMO} systems},
  journal = J_IEEE_TSP,
  year    = {2013},
  volume  = {61},
  number  = {9},
  pages   = {2204--2218},
  month   = may
}

@article{jay18,
  author  = {P. Jayasinghe and A. T{\"o}lli and J. Kaleva and M. Latva-aho},
  title   = {Bi-directional beamformer training for dynamic {TDD} networks},
  journal = J_IEEE_TSP,
  year    = {2018},
  volume  = {66},
  number  = {23},
  pages   = {6252--6267},
  month   = dec
}

@article{tol19,
  author  = {A. T{\"o}lli and H. Ghauch and J. Kaleva and P. Komulainen and M. Bengtsson and M. Skoglund and M. Honig and E. Lahetkangas and E. Tiirola and K. Pajukoski},
  title   = {Distributed coordinated transmission with forward-backward training for {5G} radio access},
  journal = M_IEEE_CM,
  year    = {2019},
  volume  = {57},
  number  = {1},
  pages   = {58--64},
  month   = jan
}

@ARTICLE{Shi14,
  author  = {C. Shi and R. A. Berry and M. L. Honig},
  title   = {Bi-directional training for adaptive beamforming and power control in interference networks},
  journal = J_IEEE_TSP,
  year    = {2014},
  volume  = {62},
  number  = {3},
  pages   = {607--618},
  month   = feb
}

@article{Ngo17,
  author  = {H. Q. Ngo and A. Ashikhmin and H. Yang and E. G. Larsson and T. L. Marzetta},
  title   = {Cell-free massive {MIMO} versus small cells},
  journal = J_IEEE_TWC,
  year    = {2017},
  volume  = {16},
  number  = {3},
  pages   = {1834--1850},
  month   = mar
}

@ARTICLE{Atz21,
  author  = {I. Atzeni and B. Gouda and A. T{\"o}lli},
  title   = {Distributed precoding design via over-the-air signaling for cell-free massive {MIMO}},
  journal = J_IEEE_TWC,
  year    = {2021},
  volume  = {20},
  number  = {2},
  pages   = {1201--1216},
  month   = feb
}

@ARTICLE{Gou24,
  author  = {B. Gouda and I. Atzeni and A. T{\"o}lli},
  title   = {Pilot-aided distributed multi-group multicast precoding design for cell-free massive {MIMO}},
  journal = J_IEEE_TWC,
  year    = {2024},
  volume  = {23},
  number  = {8},
  pages   = {9282--9298},
  month   = aug
}

@article{gou24uldl,
  author  = {B. Gouda and A. Arvola and I. Atzeni and A. T{\"o}lli},
  title   = {Combined {DL}-{UL} distributed beamforming design for cell-free massive {MIMO}},
  journal = L_IEEE_WCL,
  year    = {2024},
  volume  = {13},
  number  = {6},
  pages   = {1621--1625},
  month   = jun
}

@article{Love08overview,
  author  = {D. J. Love and R. W. Heath and V. K. N. Lau and D. Gesbert and B. D. Rao and M. Andrews},
  title   = {An overview of limited feedback in wireless communication systems},
  journal = J_IEEE_JSAC,
  year    = {2008},
  volume  = {26},
  number  = {8},
  pages   = {1341--1365},
  month   = oct
}

@article{Jindal06TIT,
  author  = {N. Jindal},
  title   = {{MIMO} broadcast channels with finite-rate feedback},
  journal = J_IEEE_TIT,
  year    = {2006},
  volume  = {52},
  number  = {11},
  pages   = {5045--5060},
  month   = nov
}

@ARTICLE{Mon06,
  author  = {B. Mondal and R. W. Heath},
  title   = {Channel adaptive quantization for limited feedback {MIMO} beamforming systems},
  journal = J_IEEE_TSP,
  year    = {2006},
  volume  = {54},
  number  = {12},
  pages   = {4717--4729},
  month   = dec
}

@article{Love04ValueFeedback,
  author  = {D. J. Love and R. W. Heath and W. Santipach and M. L. Honig},
  title   = {What is the value of limited feedback for {MIMO} channels?},
  journal = M_IEEE_CM,
  year    = {2004},
  volume  = {42},
  number  = {10},
  pages   = {54--59},
  month   = oct
}

@ARTICLE{Gao15CSIT,
  author={Gao, Zhen and Dai, Linglong and Han, Shuangfeng and I, Chih-Lin and Wang, Zhaocheng and Hanzo, Lajos},
  title   = {Compressive sensing techniques for next-generation wireless communications},
  journal = J_IEEE_WC,
  year    = {2018},
  volume  = {25},
  number  = {3},
  pages   = {144--153},
  month   = jun
}

@ARTICLE{Wen18DeepCSI,
  author  = {C.-K. Wen and W.-T. Shih and S. Jin},
  title   = {Deep learning for massive {MIMO} {CSI} feedback},
  journal = L_IEEE_WCL,
  year    = {2018},
  volume  = {7},
  number  = {5},
  pages   = {748--751},
  month   = oct
}

@ARTICLE{Guo20DLCSI,
  author  = {J. Guo and C.-K. Wen and S. Jin and G. Li},
  title   = {Convolutional neural network-based multiple-rate compressive sensing for massive {MIMO} {CSI} feedback: Design, Simulation, and Analysis},
  journal = J_IEEE_TWC,
  year    = {2020},
  volume  = {19},
  number  = {4},
  pages   = {2827--2840},
  month   = Apr
}

@ARTICLE{CsiNetLSTM,
 author={Guo, Jianhua and Wang, Lei and Li, Feng and Xue, Jiang},
  title   = {{CSI} feedback with model-driven deep learning of massive {MIMO} systems},
  journal = L_IEEE_CL,
  year={2022},
  volume={26},
  number={3},
  pages={547--551},
  month   = mar
}

@ARTICLE{Song22TemporalCSI,
  author={Li, Xiangyi and Wu, Huaming},
  journal=L_IEEE_WCL, 
  title={Spatio-Temporal Representation With Deep Neural Recurrent Network in {MIMO} {CSI} Feedback}, 
  year={2020},
  volume={9},
  number={5},
  pages={653--657},
  month = may
  }

@ARTICLE{Ma25,
  author={Ma, Yifan and He, Hengtao and Song, Shenghui and Zhang, Jun and Letaief, Khaled B.},
  journal=J_IEEE_TWC, 
  title={Low-Complexity {CSI} Feedback for {FDD} Massive {MIMO} Systems via Learning to Optimize}, 
  year={2025},
  volume={24},
  number={4},
  pages={3483--3498},
  month = apr
  }

@INPROCEEDINGS{Ju23,
  author={Ju, Hyungyu and Jeong, Seokhyun and Kim, Seungnyun and Shim, Byonghyo},
  booktitle=C_IEEE_ICC, 
  title={Transformer-Aided Parametric {CSI} Feedback for mmWave Massive {MIMO} Systems}, 
  year={2023},
  volume={},
  number={},
  pages={3596-3601},
  month=may
}

@ARTICLE{ning26,
  author  = {B. Ning and H. Yin and S. Liu and H. Deng and S. Yang and Y. Zhang and W. Mei and D. Gesbert and J. Park and R. W. Heath and E. Bj{\"o}rnson},
  title   = {Precoding Matrix Indicator in the {5G NR} Protocol: A Tutorial on {3GPP} Beamforming Codebooks},
  journal = J_IEEE_CST,
  year    = {2026},
  volume  = {28},
  number  = {},
  pages   = {4581--4623},
  month   = jan
}

@TechReport{TS38214,
  author      = "{3GPP}",
  title       = {{NR}; Physical Layer Procedures for Data ({Release}~18)},
  institution = "{3GPP}",
  year        = {2025},
  number      = {TS 38.214 V18.7.0 (2025-06)}
}

@article{fu23,
  author  = {X. Fu and D. Le Ruyet and R. Visoz and V. Ramireddy and M. Grossmann and M. Landmann and W. Quiroga},
  title   = {A tutorial on downlink precoder selection strategies for {3GPP} {MIMO} codebooks},
  journal = J_IEEE_ACCESS,
  year    = {2023},
  volume  = {11},
  pages   = {138897--138922},
  month = dec
}

@article{Gesbert10JSAC,
  author  = {D. Gesbert and S. Hanly and H. Huang and S.S Shamai and O. Simeone and W. Yu},
  title   = {Multi-cell {MIMO} cooperative networks: A new look at interference},
  journal = J_IEEE_JSAC,
  year    = {2010},
  volume  = {28},
  number  = {9},
  pages   = {1380--1408},
  month   = dec
}

@article{Irmer11CoMP,
  author  = {R. Irmer and H. Droste and P. Marsch and M. Grieger and G. Fettweis and S. Brueck and H.-P. Mayer and E. Thiele and V. Jungnickel},
  title   = {Coordinated multipoint: Concepts, performance, and field trial results},
  journal = M_IEEE_CM,
  year    = {2011},
  volume  = {49},
  number  = {2},
  pages   = {102--111},
  month   = feb
}

@article{Bjornson12CombiningVsMultiplexing,
  author  = {E. Bj{\"o}rnson and M. Kountouris and M. Bengtsson and B. Ottersten},
  title   = {Receive combining vs. multi-stream multiplexing in downlink systems with multi-antenna users},
  journal = J_IEEE_TSP,
  year    = {2013},
  volume  = {61},
  number  = {13},
  pages   = {3431--3446},
  month   = jul
}

@article{Tru21,
author = {Trung Vu and Evgenia Chunikhina and Raviv Raich},
title = {Perturbation expansions and error bounds for the truncated singular value decomposition},
journal = {Linear Algebra and its Applications},
volume = {627},
pages = {94--139},
year = {2021},
month = oct
}

@article{Scu14,
	author = {Scutari, G. and Facchinei, F. and Song, P. and Palomar, D. P. and {Pang}, J.-S.}, 
	journal = J_IEEE_TSP,
	title = {Decomposition by Partial Linearization: Parallel Optimization of Multi-Agent Systems},
	year = {2014},
	volume = {62},
	number = {3},
	pages ={641--656},
	month = {Feb.}}

@article{Vie17Reciprocity,
  author  = {J. Vieira and F. Rusek and O. Edfors and S. Malkowsky and L. Liu and F. Tufvesson},
  title   = {Reciprocity calibration for massive {MIMO}: Proposal, modeling, and validation},
  journal = J_IEEE_TWC,
  year    = {2017},
  volume  = {16},
  number  = {5},
  pages   = {3042--3056},
  month   = may,
  doi     = {10.1109/TWC.2017.2674659}
}

@ARTICLE{Da00,
  author={Da-Shan Shiu and Foschini, G.J. and Gans, M.J. and Kahn, J.M.},
  journal=J_IEEE_TC, 
  title={Fading correlation and its effect on the capacity of multielement antenna systems}, 
  year={2000},
  volume={48},
  number={3},
  pages={502--513},
  month = mar
  }

@INPROCEEDINGS{Gou26,
  author={Gouda, Bikshapathi and Arvola, Antti and Karjalainen, Juha and Hakola, Sami and Tölli, Antti},
  booktitle=C_EUCNC_6GSUMMIT, 
  title={Precoding Design with Codebook-Based Effective {CSI} Feedback in {MIMO} Systems}, 
  year={2026},
  volume={},
  number={},
  pages={878-883},
 }

@ARTICLE{Her11,
  author={Herath, S. P. and Rajatheva, N. and Tellambura, C.},
  journal=J_IEEE_TC, 
  title={Energy Detection of Unknown Signals in Fading and Diversity Reception}, 
  year={2011},
  volume={59},
  number={9},
  pages={2443--2453},
  month=sep
  }

@ARTICLE{Sun10,
  author={Sun, Yin and Baricz, {\'A} and Zhou, Shidong},
  journal=J_IEEE_TIT, 
  title={On the Monotonicity, Log-Concavity, and Tight Bounds of the Generalized Marcum and Nuttall  {Q}-Functions}, 
  year={2010},
  volume={56},
  number={3},
  pages={1166--1186},
  month=mar
  }

@ARTICLE{Kal18,
  author={Kaleva, Jarkko and Tölli, Antti and Juntti, Markku and Berry, Randall A. and Honig, Michael L.},
  journal=J_IEEE_TSP, 
  title={Decentralized Joint Precoding With Pilot-Aided Beamformer Estimation}, 
  year={2018},
  volume={66},
  number={9},
  pages={2330-2341},
  month=may
  }
\end{document}